\documentclass[aps,prd,a4paper,preprintnumbers,tightenlines,superscriptaddress,
twocolumn,floatfix,showpacs,final]{revtex4-1}
\usepackage{graphicx}
\usepackage{longtable}
\usepackage{amsmath}
\usepackage{amssymb}
\usepackage{subfigure}
\usepackage{hyperref}
\usepackage{dcolumn}
\usepackage{bm}
\usepackage{color}

\newcommand{\mpl}{{M_{\rm {pl}}}}

\newcommand{\dd}{\, {\rm d}}

\newcommand{\pn}{\Phi_{\rm N}}

\newcommand*\colvec[3][]{\begin{pmatrix}\ifx\relax#1\relax\else#1\\\fi#2\\#3\end{pmatrix}}
\newcommand{\tg}{\tilde{g}}
\newcommand{\iii}{_{\rm i}}
\newcommand{\nm}{{\mu\nu}}

\newcommand{\lt}{\tilde{\Lambda}}
\newcommand{\mmm}{{_{\rm m}}}
\newcommand{\rmm}{{\rho_{\rm m}}}
\newcommand{\inff}{_\infty}
\newcommand{\phii}{\phi_\infty}
\def\eea{\end{eqnarray}}
\def\bea{\begin{eqnarray}}
\allowdisplaybreaks
\usepackage{enumitem}
\setenumerate[1]{label=(\arabic*)}
\begin{document}
\title{Towards Viable Cosmological Models of Disformal Theories of Gravity}
\author{Jeremy Sakstein}
\affiliation{Department of Applied Mathematics and Theoretical Physics, Centre
for Mathematical Sciences, Cambridge CB3 0WA,
United Kingdom}
\affiliation{Institute of Cosmology and Gravitation,
University of Portsmouth, Portsmouth PO1 3FX, UK}
\affiliation{Perimeter Institute for Theoretical Physics, 31 Caroline St. N,
Waterloo, ON, N2L 6B9, Canada}

\begin{abstract}

The late-time cosmological dynamics of disformal gravity are investigated using dynamical systems methods. It is shown that in the general case 
there are no stable attractors that screen fifth-forces locally and simultaneously describe a dark energy dominated universe. Viable scenarios have 
late-time properties that are independent of the disformal parameters and are identical to the equivalent conformal quintessence model. Our analysis 
reveals that configurations where the Jordan 
frame metric becomes singular are only reached in the infinite future, thus explaining the natural pathology resistance observed numerically by 
several previous works. The viability of models where this can happen is discussed in terms of both the cosmological dynamics and local phenomena. We 
identify a special parameter tuning such that there is a new fixed point that can match the presently observed dark energy 
density and equation of state. This model is unviable when the scalar couples to the visible sector but may provide a good candidate model for 
theories where only dark matter is disformally coupled. 
\end{abstract}
\maketitle

\section{Introduction}

The observation of the acceleration of the cosmic expansion \cite{Riess:1998cb,Perlmutter:1998np} has prompted a renewed interest in modified 
theories of gravity as a potential driving mechanism. Amongst the plethora of candidate theories (see \cite{Clifton:2011jh} for a review), those 
that contain screening mechanisms (see \cite{Joyce:2014kja} for a review) have been particularly well-studied due to their ability to hide the 
additional or \textit{fifth}- forces on solar system scales. Well-studied non-linear screening mechanisms---meaning that the local dynamics act to 
suppress fifth-forces---include the chameleon mechanism \cite{Khoury:2003aq,Khoury:2003rn} and similar \cite{Hinterbichler:2011ca,Brax:2010gi} as 
well 
as the Vainshtein mechanism \cite{Vainshtein:1972sx}. These theories can be described by a conformal coupling of a scalar to matter through the 
metric\footnote{Note that some theories such as massive gravity \cite{deRham:2010kj} utilise these mechanisms, although the coupling to matter is 
only seen explicitly in the decoupling limit.}. Theories that contain a disformal coupling to matter have been studied in the context of dark matter 
and dark energy 
\cite{1992mgm..conf..905B,Bekenstein:1992pj,Koivisto:2008ak,Zumalacarregui:2010wj,Noller:2012sv,Zumalacarregui:2012us,Zumalacarregui:2013pma,
Koivisto:2013fta,Koivisto:2014gia,Sakstein:2014isa} (see also \cite{Kaloper:2003yf} for an application to 
inflation) but unlike chameleons, which have been well studied and constrained on small scales 
\cite{Davis:2011qf,Jain:2012tn,Brax:2013uh,Sakstein:2013pda,Vikram:2014uza,Sakstein:2014nfa}, the local behaviour of disformal theories has been 
relatively unstudied 
(although see \cite{Freire:2012mg,Khoury:2014tka} for tests of TeVeS and a model that satisfies the Cassini bound on light bending by the Sun). 

Recently, \cite{Sakstein:2014isa} has performed a thorough investigation into the local dynamics of the scalar. Non-relativistic 
objects inside a Friedmann-Robertson-Walker (FRW) universe will source a field profile such that the fifth-force is $F_5=2Q^2F_{\rm N}$---$F_{\rm N}$ 
is the Newtonian force---where the local scalar charge $Q$ depends on the conformal coupling, the disformal parameters and the first and second 
time-derivatives of the cosmological field. There, it was argued that there are no non-linear screening mechanisms beyond those mentioned above. It 
was shown, however, that when the conformal coupling is absent\footnote{One could screen this using the 
Damour-Polyakov effect \cite{Damour:1994zq}, although this has not yet been investigated.} these theories screen linearly---by which we mean 
the local scalar charge is suppressed on all scales through the cosmological dynamics and not the local ones---whenever the cosmological dynamics are 
such that the 
field is slowly rolling.

The aim of this work is to investigate more general models where the conformal factor is non-zero and see whether one can find models that 
can drive the cosmic acceleration at late times whilst simultaneously suppressing the local scalar charge. We will do this using dynamical systems 
techniques. These are powerful tools to classify the late-time behaviour of non-linear systems despite the lack of analytic solutions. Their power 
lies in their ability to classify all solutions of the system in terms of a few fixed points independently of the initial conditions. Without them, 
one would be reduced to solving the problem numerically for all possible parameters and initial conditions, a problem that is clearly intractable. 
Since the relevant cosmological parameters have well-defined values at these points, one can know the final state the universe will evolve to. 
Dynamical systems methods have previously been used in the study of dark energy models in the form of quintessence \cite{Copeland:1997et} and 
conformally coupled theories \cite{Holden:1999hm,Gumjudpai:2005ry} (see \cite{Copeland:2006wr} for a review). There, they find dark energy dominated 
solutions by 
looking for points where the density parameter and equation of state is close to the observed values \cite{Hinshaw:2012aka}. Here, we will do the 
same for the disformal system, examining the local scalar charge as well in order to classify the linear screening. Unlike the simple cases mentioned 
above, the disformal system requires use of the centre manifold technique, which is necessary whenever different directions in phase space evolve on 
different time-scales. Interestingly, we will see that some of the attractors of the conformal system are saddle points of the disformal system and 
the evolution can be very different at late times.

These theories have potential instabilities since the determinant of the Jordan frame metric can become singular. Several authors 
\cite{Koivisto:2008ak,Zumalacarregui:2010wj,Zumalacarregui:2012us} have observed numerically that solutions tend to slow down to avoid this and have 
dubbed this phenomena a \textit{natural resistance to pathology}. Using the techniques described above, we will show that any singularity is only 
reached in an infinite amount of time, thereby explaining this phenomena. Whether or not this is a pathology of the theory has been debated and here 
we will argue that whenever this singularity is present, the non-relativistic limit is not well defined and one generally expects large fifth-forces 
and time-variations in Newton's constant as the singularity is approached. 

%
%
Since portions of this paper are technical, we state our main results unambiguously below:
\begin{itemize}
 \item The phase space is three-dimensional, in contrast to the purely conformal case. In the general case, there is one new dark energy dominated 
solution (discussed in the next bullet point) and one retains all of the solutions found in the purely conformal case. The dark energy dominated 
points found also in the conformal case do not screen linearly unless the conformal factor is absent.
\item There are parameter choices where the purely conformal fixed points are saddle points. In this case, the system ultimately evolves towards a 
dark energy dominated solution where the determinant of the Jordan frame metric is singular and there are large unscreened fifth-forces.
\item There is a special tuning in parameter space where the dimension of the phase space is reduced to two. In this case there is a new stable 
attractor where one can reproduce the present dark energy density parameter and equation of state by making an appropriate choice for the model 
parameters. 
Unfortunately, the metric singularity is approached along this attractor and hence the model is not viable.
\item The conclusions drawn here assume that the disformal coupling of the scalar to matter is universal. If one were to relax this and only couple 
the scalar to the dark sector (i.e. visible matter couples only to the Einstein frame metric) then many of the solutions claimed to be unviable here 
may be tenable. This requires further investigation and we discuss this in detail in the conclusions.
\end{itemize}

This paper is organised as follows: The theory is introduced in section \ref{sec:disf} where we discuss the local fifth-forces, the metric 
singularity and the potential problems it entails. Section \ref{sec:dyn} constitutes the main body of this work. 
There it is shown that there are no stable dark energy dominated solutions that simultaneously have zero local scalar charge. Furthermore, we show 
that the singularity in the Jordan frame metric is only reached in the infinite future. In section \ref{sec:red} we examine the special parameter 
choices such that the phase space is two-dimensional and find a new fixed point that can reproduce the observed dark energy 
parameters but has a metric singularity in the infinite future. We conclude in section \ref{sec:concs}. A brief introduction to dynamical systems 
including centre manifold techniques is provided in Appendix \ref{sec:dyna} for the unfamiliar reader.

\section{Disformal Theories of Gravity}\label{sec:disf}

Disformal theories of gravity are described by the following action:
\begin{equation}\label{eq:act}
 S=\mpl^2\int\dd^4x\sqrt{-g}\left[\frac{R}{2}+X-V(\phi)\right]+S_{\rm
m}\left[\tg;\Psi\iii\right],
\end{equation}
where the various matter fields $\Psi\iii$ are coupled to the Jordan frame metric
\begin{equation}\label{eq:metric}
 \tg_{\mu\nu}=A^2(\phi)\left(g_{\mu\nu}+\frac{B^2(\phi)}{\Lambda^2}\partial_\mu\phi\partial_\nu\phi\right),
\end{equation}
and $X\equiv -1/2\nabla_\mu\phi\nabla^\mu\phi$. $g_\nm$ is the Einstein frame metric. $A$ and $B$ are known as the \textit{conformal} and 
\textit{disformal}
factors respectively and we define the following quantities\footnote{Note that we are using the conventions of \cite{Sakstein:2014isa}. A dictionary 
to convert these conventions to others in the literature such as \cite{Zumalacarregui:2012us} can be found there.}
\begin{equation}\label{eq:agdefs}
 \alpha(\phi)\equiv\frac{\dd\ln A(\phi)}{\dd \phi}\quad \textrm{and}\quad\gamma(\phi)\equiv\frac{\dd\ln B(\phi)}{\dd \phi}.
\end{equation}
Varying the action (\ref{eq:act}) with respect to $\phi$, one finds the equation of motion for the field
\begin{equation}\label{eq:QQ}
 \Box\phi=V(\phi)_{,\,\phi}+Q,
\end{equation}
where
\begin{widetext}
\begin{equation}\label{eq:Qtmunudef}
 Q\equiv
\nabla_\mu\left(\frac{B^2(\phi)}{\Lambda^2}T\mmm^{\mu\nu}\nabla_\mu\phi\right)-\alpha(\phi)T\mmm-\frac{B(\phi)^2}{\Lambda^2}
\left[\gamma(\phi)+\alpha(\phi)\right]T\mmm^{\mu\nu}\nabla_\mu\phi\nabla_\nu\phi.
\end{equation}
\end{widetext}
Here $T_m^\nm=2/\sqrt{-g}\delta S\mmm/\delta g_\nm$ is the energy-momentum tensor for matter. Since the scalar is coupled directly to matter, this is 
not conserved\footnote{It is the Jordan frame energy-momentum tensor that is conserved since matter is minimally coupled in this frame.} and one 
instead has
\begin{equation}
 \nabla_\mu T\mmm^{\mu\nu}=-Q\nabla^\nu\phi.
\end{equation}
Contracting this with $\nabla_\nu\phi$, one can algebraically solve for $\nabla_\mu T^\nm\nabla_\nu\phi$, which can be used to eliminate all 
derivatives of the energy-momentum tensor in (\ref{eq:Qtmunudef}) so that (\ref{eq:QQ}) becomes
\begin{widetext}
\begin{align}
\left(1-\frac{2B^2X}{\Lambda^2}\right)\Box\phi&-8\pi G\frac{B^2}{\Lambda^2}T_{\rm m}^{\mu\nu}\nabla_\mu\nabla_\nu\phi=-8\pi\alpha
GT_{\rm m}-8\pi G\frac{B^2}{\Lambda^2}\left(\alpha-\gamma\right)T_{\rm
m}^{\mu\nu}\partial_\mu\phi\partial_\nu\phi+\chi V(\phi)_{\phi}.\label{eq:sfeom}
\end{align}
\end{widetext}
In practice, this equation is the more useful of the two work with and here we will do so whenever possible.

The determinants of the two metrics are related by (see \cite{Zumalacarregui:2012us}, Appendix A)
\begin{equation}\label{eq:metrel}
 \frac{\sqrt{-\tg}}{\sqrt{-g}}=A^4\sqrt{1-\frac{2B^2X}{\Lambda^2}}.
\end{equation}
Note that it is possible for the right hand side to become zero so that the metric is singular, which is a potential problem for the theory. 
Cosmologically, several authors \cite{Koivisto:2008ak,Zumalacarregui:2010wj,Zumalacarregui:2012us} have numerically found a \textit{natural 
resistance to pathology} in the solutions where the cosmological time-evolution of the field slows down at late times and the singularity is avoided. 
We will address this issue 
later and show that certain models can evolve towards the metric singularity but only in an infinite time, thereby explaining the numerical 
observations of this pathology resistance.

\subsection{Local Behaviour and Pathologies}\label{sec:via}

In \cite{Sakstein:2014isa}, it was shown that when expanded around an FRW background, the fifth-force arising in these theories is 
\begin{equation}\label{eq:f5}
 F_5=2Q^2F_{\rm N},
\end{equation}
where $F_{\rm N}$ is the Newtonian force and the local scalar charge $Q$ is 
\begin{equation}\label{eq:Qdef}
 Q\equiv \frac{\alpha+\frac{B^2}{\Lambda^2}\left(\ddot{\phi}_\infty+\dot{\phi}_\infty^2\left[\gamma-\alpha\right] 
\right)}{1-\frac{B^2\dot{\phi}^2_\infty}{\Lambda^2}},
\end{equation} 
where $\phi\inff$ is the cosmological (homogeneous) component of the field and an over-dot denotes a derivative with respect to coordinate time $t$ 
in the Einstein frame. For completeness, we briefly present the derivation of this formula here. Using the Einstein frame coordinates $\dd 
s^2=-(1+2\Phi)\dd t^2+ (1-2\Psi)\delta_{ij}\dd x^i\dd x^j$ we expand the field equations about the cosmological background value so that 
$\phi-\phi_\infty(t)+\varphi(r,t)$ and ignore all terms that are post-Newtonian\footnote{It is assumed that $\Phi\sim\Psi\sim\varphi\sim\pn$, where 
$\pn$ is the Newtonian potential, so that terms such as $\dot{\varphi}^2\sim\pn\nabla^2\pn$ can be neglected. This ansatz can be checked 
self-consistently (see the discussion in \cite{Sakstein:2014isa}) and allows one to systematically construct the non-relativistic limit.}. With these 
assumptions, equation (\ref{eq:sfeom}) becomes
\begin{equation}
\nabla^2 \varphi= 8\pi QG\rho.
\end{equation}
This is nothing but the Poisson equation with $G\rightarrow 2Q G$ so that $\varphi=2 Q\pn$, where $\pn$ is the Newtonian potential. Next, we need the fifth-force, which can be found as follows. Defining the tensor 
\begin{equation}
\mathcal{K}^\alpha_\nm\equiv \tilde{\Gamma}^\alpha_\nm-\Gamma_\nm^\alpha,
\end{equation}
the geodesic equation can be written (note that matter moves on geodesics of $\tg_\nm$ since this is the metric that couples to matter) in terms of purely Einstein frame quantities. In the non-relativistic limit one has $\dd x^0/\dd \tau\gg\dd x^i/\dd\tau$---$\tau$ is the proper-time in the Jordan frame---so that 
\begin{equation}
\ddot{x}^i+\Gamma^i_{00}=-\mathcal{K}_{00}^i.
\end{equation}
In this form, one can use the familiar result $\Gamma^i_{00}=\partial^i\pn$\footnote{Note that this assumes $\dot{\phi}_\infty\ll\Lambda$. If this is 
not the case then there are time-dependent corrections to Newton's constant (see the discussion below).} to see that all of the effects of the 
fifth-force are contained within $\mathcal{K}^i_{00}$. Using equation (\ref{eq:metric}), one finds that $\mathcal{K}^i_{00}=Q\partial^i\varphi$ (see 
\cite{Sakstein:2014isa}) so that the fifth-force is given by $F_5=2Q^2F_{\rm N}$, where $F_{\rm N}$ is the Newtonian force. The theory then behaves as 
one with a time-varying gravitational constant $G_{\rm eff}=G(1+2Q^2)$ when $\dot{\phi}_\infty\ll\Lambda$ (see the discussion below).

In \cite{Sakstein:2014isa}, we examined models where $\alpha(\phi)=0$ and
argued 
that any model where the scalar potential has a minimum can naturally screen fifth-forces because at late-times, when the field is slowly-rolling, 
$Q\approx0$. In this work, we wish to address the more general question: can we find models with $\alpha\ne0$ where the scalar charge is identically 
zero at late times and the universe is dark energy dominated? In order to answer this, one must use dynamical systems techniques since this allows 
one to explore the entire solution space of the theory without numerically solving each model with every possible set of initial conditions. 

One important question to address is whether solutions that evolve towards a singularity in the infinite future should be considered viable or not. 
In this subsection we will argue that they should not, at least when the scalar couples to the visible sector.

Note that the denominator in (\ref{eq:Qdef}) is zero precisely when the metric singularity occurs. The simplest interpretation of this is that 
the fifth-force becomes infinite whenever the numerator does not tend to zero at the same rate. In this case one would denounce these 
models as unviable unless the numerator approaches zero. Such a property can not be a generic feature of any given model. We will see below that the 
phase space of cosmological solutions is three-dimensional and so the lack of a multi-dimensional analogue of L' H\^{o}pital's rule means that the 
value of $Q$ is not set by the properties of the fixed point. Instead, its value depends on the trajectory through phase space along which the point 
is approached. For this reason, a solution where fifth-forces are screened at late times using this method is only found using a special tuning of 
both the model parameters and the initial conditions. 

The discussion above is predicated on the assumptions that the non-relativistic limit derived in \cite{Sakstein:2014isa} is valid 
when $B\dot{\phi}_\infty\rightarrow\Lambda$ and that one can use the Einstein frame coordinates to describe the physics of local observers. In fact, 
when this is the case the non-relativistic limit is far more subtle and may not even exist. To see this, consider the case where 
$A(\phi)=B(\phi)=1$\footnote{One could perform the same analysis for the general model but this is cumbersome and does not add any more insight. The 
pathologies discussed here arise due to the disformal coupling and so it suffices to consider the simplest case for the purposes of elucidation.}. If 
one chooses coordinates $\{t,x^i\}$ such that the Einstein frame metric is FRW, i.e. $g_{\mu\nu}=\mathrm{diag}(-1,a_{\rm E}^2(t),a_{\rm 
E}^2(t),a_{\rm E}^2(t))$ and the homogeneous component of the scalar is $\phi_\infty(t)$ then using equation (\ref{eq:metric}) one finds that the 
line-element in the Jordan frame is
\begin{equation}\label{eq:lejf1}
 \dd\tilde{s}=-\mathcal{N}^2\dd t^2+a_{\rm E}^2(t)\delta_{ij}\dd x^i\dd x^j, 
\end{equation}
where the Jordan frame lapse is 
\begin{equation}\label{eq:Ndef}
\mathcal{N}^2\equiv 1-\frac{\dot{\phi}_\infty^2}{\Lambda^2}. 
\end{equation}
One can immediately see the Jordan frame manifestation of the metric singularity: the lapse becomes zero. When $\dot{\phi}_\infty\sim\Lambda$ 
the proper time for stationary observers in the Jordan frame is not given by $t$ but rather by $T$, where
\begin{equation}\label{eq:T}
 \dd T= \mathcal{N}\dd t.
\end{equation}
Since it is the Jordan frame metric that couples directly to matter, it is this metric that one must use to calculate the observables in this theory. 
Indeed, the non-relativistic limit was derived in \cite{Sakstein:2014isa} by transforming the non-relativistic limit of the geodesic equation to the 
Einstein frame in order to allow an interpretation of the new effects as a fifth-force. This derivation assumes a locally-inertial reference frame, 
which requires one to work with $T$ and not $t$. Thus, when $\dot{\phi}_\infty\ll\Lambda$ this is perfectly valid and equation (\ref{eq:f5}) 
describes a fifth-force due to a field with scalar charge
\begin{equation}\label{eq:Qdef2}
 Q\approx\alpha+\frac{B^2}{\Lambda^2}\left(\ddot{\phi}_\infty+\dot{\phi}_\infty^2\left[\gamma-\alpha\right] 
\right).
\end{equation} 
When the converse is true, the non-relativistic limit, if it exists, must be found using $T$ and not $t$. Defining a non-relativistic limit in this 
limit is difficult because one actually has two speeds in the problem: the speed of light, $c_{\rm light}=c$, and the speed of tensor propagation in 
the Jordan frame (see \cite{Bellini:2014fua})
\begin{equation}
 c_{\rm tensors}^2=1-\frac{\dot{\phi}_\infty^2}{\Lambda^2}.
\end{equation}
Note that one can instead interpret this as a varying speed of light in the Einstein frame \cite{vandeBruck:2012vq,Brax:2013nsa,vandeBruck:2013yxa}, 
although one must be careful to interpret this speed correctly, especially when different species couple to different metrics. It is not 
clear whether a consistent non-relativistic limit exists when these differ greatly since one must choose a small parameter in order to expand the 
equations. This is typically $v/c$ where $v$ is a typical speed scale in the problem but a priori one does not know whether this should be $v/c_{\rm 
light}$, $v/c_{\rm tensors}$ or some combination of the two. Since the speed of tensors becomes very low when $\dot{\phi}_\infty\sim\Lambda$ one may 
worry that there is no non-relativistic limit---or rather, that it is valid at speeds far smaller than those present in the solar system---since one 
of the speeds in the problem is very low. A full analysis of this would require a complete self-consistent solution of the field equations in the 
Jordan frame to determine which, if any, approximations can be made. This is clearly well beyond the scope of this work. 

Other potential issues that arise when $\dot{\phi}_\infty\sim\Lambda$ can be seen examining the Jordan frame action, which contains a term of the 
form (this can be found using the method of \cite{Bettoni:2013diz}):
\begin{equation}
 S\supset\int\dd^4 x\sqrt{-\tg}\frac{R}{16\pi G }\sqrt{1-\frac{\tilde{\nabla}_\mu\phi\tilde{\nabla}^\mu\phi}{\Lambda^2}},
\end{equation}
where 
\begin{equation}
 \tilde{\nabla}_\mu\phi\tilde{\nabla}^\mu\phi=\tg^{\nm}\partial_\mu\phi\partial_\nu\phi.
\end{equation}
Using the coordinate system defined by (\ref{eq:T}), one finds:
\begin{equation}
 \frac{\dd \phi_\infty}{\dd T}=\frac{\dot{\phi}_\infty}{\mathcal{N}}
\end{equation}
and
\begin{equation}
 \dd\tilde{s}=-\dd T^2+a_{\rm J}^2(T)\delta_{ij}\dd x^i\dd x^j, 
\end{equation}
where the Jordan frame scale factor is $a_{\rm J}(T)=a_{\rm E}(t(T))$. From this, one can see that the effective value of Newton's constant in the 
solar system is
\begin{equation}
G_{\rm N}(t)= G\sqrt{1-\frac{\dot{\phi}^2_\infty}{\Lambda^2}}\left[1+f(\partial_i\varphi)\right],
\end{equation}
where $\phi=\phi_\infty+\varphi$ and $f(\partial_i\varphi)$ represents the effects of the inhomogeneous component of the field sourced by small-scale 
sources. One can see that theories where $\dot{\phi}_\infty$ approaches $\Lambda$ predict lower values of $G_{\rm N}$ in the solar system than would 
be 
inferred using the Friedmann equation (this is similar to conformal theories where $G_{\rm N}=A^2(\phi_\infty)G$ \cite{Brax:2012yi}). This behaviour 
is also seen in other theories that include non-minimal couplings to curvature tensors \cite{Barreira:2013xea}.

Finally, note that knowledge of the Einstein frame dynamics is not sufficient to determine the late-time dynamics in the Jordan frame when the 
solution approaches the metric singularity. To see this, note that
\begin{equation}
 H_{\rm J}=\frac{H_{\rm E}}{\mathcal{N}}\quad\textrm{and}\quad \frac{1}{a_{\rm J}}\frac{\dd^2 a_{\rm J}}{\dd 
T^2}=\frac{1}{\mathcal{N}^2}\left(\frac{\ddot{a}_{\rm E}}{a_{\rm E}}+H_{\rm E}\frac{\ddot{\phi}_\infty}{\Lambda^2}\right).
\end{equation}
At late times, we expect $H_{\rm E}\rightarrow0$ but since $\mathcal{N}\rightarrow0$ in the same limit the behaviour of $H_{\rm J}$ cannot be 
determined using dynamical systems methods\footnote{This is because the analysis employed below uses $\ln a(t)$ as a time coordinate and eliminates 
$H(a)$ through the Friedmann constraint. This choice implicitly assumes that $a(t)$ is monotonically increasing and can hence be used as a time 
coordinate. One could instead work with a larger phase space in order to treat $H$ as a dynamical variable and capture any deviations from 
monotonicity but it is simpler to work in the Jordan frame. This exercise is postponed for follow-up work.}. As far as the acceleration of the 
universe is concerned, one expects the Einstein frame Hubble parameter to approach zero at late times and so an Einstein frame analysis is 
sufficient to determine whether or not $\ddot{a}_{\rm J}>0$, which is the true criterion for acceleration\footnote{One may wonder whether 
re-introducing $A$ and $B$ can change this argument. One can always perform a field re-definition to remove the function multiplying the disformal 
factor and hence any effects of having $B\ne1$ can be equivalently thought of as arising due to non-canonical kinetic terms for the scalar. 
The conformal factor is less subtle but in this case any acceleration in the Jordan frame is due to conformal effects only and hence this scenario 
is not relevant for this work.}.

It is for the reasons presented in this subsection that we will consider models that approach the metric singularity as unviable in this work. We 
will discuss possible caveats and resolutions in the conclusions as well as possible future directions one could pursue in order to address these 
issues further.

\subsection{Cosmology}

From here on we will work exclusively in the Einstein frame and we hence drop any subscripts on cosmological quantities such as $H$ and $a$. All 
quantities are given in the Einstein frame unless explicitly stated otherwise. The background cosmology (in coordinate time) is determined by the 
Friedmann equations
\begin{align}
 3H^2&= 8\pi G\rho\mmm+\frac{\dot{\phi}_\infty^2}{2}+V(\phii)\label{eq:fried1}\\
\dot{H}&=-4\pi G\rho\mmm -\frac{\dot{\phi}_\infty^2}{2}\label{eq:fried2}
\end{align}
coupled to the equation of motion for the scalar and the matter density:
\begin{align}
 \ddot{\phi}_\infty+3H\dot{\phi}_\infty+V(\phii)_{,\phi}&=-Q_0\quad\textrm{and}\label{eq:fired3}\\
\dot{\rho}\mmm+3H\rho\mmm=Q_0\dot{\phi}_\infty\label{eq:cont}
\end{align}
 where
 \begin{equation}\label{eq:Q0}
  Q_0=8\pi G\rho\mmm\frac{\alpha+\frac{B^2}{\Lambda^2}\left(\left[\gamma-\alpha\right]
 \dot{\phi}_\infty^2-3H\dot{\phi}_\infty-V_{,\phi}\right)}{1+\frac{B^2}{\Lambda^2}\left(8\pi G
 \rmm-\dot{\phi}_\infty^2\right)},
\end{equation}
and $\rho=-T\mmm$ with $T\mmm$ the trace of the Einstein frame energy-momentum tensor. Note that since the scalar is non-minimally coupled to 
matter the energy-momentum tensor is not covariantly conserved in this frame, which leads to the modified continuity equation (\ref{eq:cont}). We do 
not include radiation or other particle species in this work since we are interested in the late-time behaviour of the system where all of these 
components are sub-dominant.

\section{Dynamical Systems Analysis}\label{sec:dyn}

\subsection{Phase Space Construction}

Dynamical systems have been used in cosmology (see \cite{Copeland:2006wr} and references therein) to examine the late time behaviour of quintessence 
\cite{Copeland:1997et} and conformally coupled dark energy \cite{Gumjudpai:2005ry}. Here, we will extend the formalism to disformally coupled 
dark energy\footnote{Note that this was studied briefly in 
\cite{Zumalacarregui:2012us}, Appendix C.}. A brief introduction to the general techniques is given in Appendix \ref{sec:dyna} for the unfamiliar 
reader. We begin by introducing 
the new variables
\bea
& &
x\equiv \frac{\phii^\prime}{\sqrt{6}}\,,~~
y \equiv \frac{\sqrt{V}}{\sqrt{3}H}\,, 
\nonumber \\
& &
\label{lamGam}
\lambda \equiv -\frac{V_{,\phi}}{V}\,,~~
z\equiv\frac{BH}{\Lambda}\,,
\eea
where we have changed from coordinate time $t$ to $N\equiv\ln a(t)$ and use a prime to denote derivatives with respect to $N$. In order to focus on 
the simplest case, we take $\lambda$, $\alpha$ and $\gamma$ to be constant so that
\begin{equation}\label{eq:model}
 V(\phi)=m_0^2e^{-\lambda\phi},\quad A(\phi)=e^{\alpha\phi}\quad\textrm{and}\quad B(\phi)=e^{\gamma\phi},
\end{equation}
where $m_0$ is a mass scale associated with the scalar potential. Note that this is the coupled dark energy model of \cite{Gumjudpai:2005ry} extended 
to include an exponential disformal coupling. Such a model has previously been studied by \cite{Zumalacarregui:2012us} with $\alpha=0$ and so this 
is a generalisation of their model to arbitrary conformal couplings. One should note that whereas this 
model makes a specific choice for the functional forms of the free functions, one can think of the fixed points found using this system as 
instantaneous fixed points for more complicated models\footnote{By this, we mean that at any fixed time the fixed points found here will be 
solutions of the equations. Whether or not they are reached however depends on the specific model and in general one expects new fixed points and 
that the properties found here may be destroyed. In practice, models where the dynamics are predictable---for example potentials with minima so that 
we expect $\lambda(\phi)\rightarrow0$ at late times---tend to approach these fixed points but we stress that a more complicated analysis is 
required to draw any definite conclusions. This is especially true in cases where either the potential or the disformal factor can become zero.}. 
Written in these variables, the Friedmann-scalar field system can be written in first-order 
autonomous form:
\begin{widetext}
\begin{align}
\frac{\dd x}{\dd N}&=-3 x+\frac{3}{2} x \left(1+x^2-y^2\right)+\sqrt{\frac{3}{2}} y^2 \lambda
\nonumber\nonumber\\\label{eq:eqx}&+\frac{\sqrt{\frac{3}{2}}
\left(x^2+y^2-1\right) \left(\alpha +3 z^2 \left(2 x^2 (\gamma-\alpha)-\sqrt{6} x+y^2 \lambda \right)\right)}{1+3 \left(1-3
x^2-y^2\right) z^2},\\ 
\frac{\dd y}{\dd N}&=\frac{3}{2} y \left(1+x^2-y^2\right)-\sqrt{\frac{3}{2}} x y
 \lambda\label{eq:eqy}\quad\textrm{and}\\ \frac{\dd z}{\dd N}&=\sqrt{6} \gamma  x z-\frac{3}{2} z
 \left(x^2-y^2+1\right).\label{eq:eqz}
 \end{align}
 \end{widetext}
These are to be supplemented with the Friedmann constraint
\begin{equation}\label{eq:fcons}
 x^2+y^2+\Omega\mmm=1,
\end{equation}
where $\Omega\mmm\equiv 8\pi G\rho\mmm/3H^2$. This system is then three-dimensional. In terms of the variables $\{\phi,a(t),\Omega\mmm\}$ we 
have a first-order equation for $\Omega\mmm$ and second-order equations for $\phi$ and $a(t)$ implying that we have a five-dimensional phase 
space. One can see that there is a symmetry of the solutions whereby we can rescale $H$ and the 
various other factors ($B$, $V$ and $\dot{\phi}_\infty$) and leave the equations invariant. For this reason, one can scale $H$ out of the problem and 
work with the five quantities $\{x,y,z,\Omega\mmm,a\}$. Equation (\ref{eq:fcons}) is a constraint equation and so we can use it to eliminate 
$\Omega\mmm$ which reduces the dimension of the phase space to four. Finally, the phase space is reduced to three by the choice of time coordinate 
$N$. Since the equations are time-translation invariant in this coordinate, this allows us to remove all dependence on $a$ in the equations so that 
$x$, $y$ and $z$ are the only dynamical degrees of freedom and thereby reducing the dimension of the phase space to three. This leaves us with three 
coupled first-order equations including the constants $\{\lambda,\alpha,\gamma\}$. This is to be contrasted with the 
conformally coupled case where the phase space is two-dimensional. This is because choosing $\lambda$ and $\alpha$ to be constant removes any 
dependence on $\phi$ in the equations for $x$ and $y$ whereas $B(\phi)$ remains in $Q_0$ even when $\gamma$ is constant. Note also that $\gamma=0$ 
does not reduce the dimensionality of the phase space because in this case $z$ depends on $H$, which cannot be eliminated in terms of $x$ and $y$ 
only and so the case where $B$ is constant is included in this analysis. On the other hand, when $\gamma=\lambda/2$ we have the relation
\begin{equation}\label{eq:ps2}
 zy=\frac{m_0}{\sqrt{3}\Lambda}
\end{equation}
and so one may eliminate $z$ in terms of $y$ or vice versa. Therefore, $\gamma=\lambda/2$ is a special hyperplane in parameter space\footnote{The 
parameter space 
is spanned by $\gamma$, $\alpha$, $\lambda$ and $\Lambda/m_0$} where the dimensionality of the phase space is reduced to two\footnote{One may solve 
for $\phi$ from the definition of $y$ to find $B=(\sqrt{3}Hy)^{2\gamma/\lambda}$. When $\gamma=\lambda/2$ one can scale $H$ out of the equations 
without the need to define a new quantity $z$ whereas when this is not the case a new variable is needed. Therefore, one may think of this tuning as 
an enhanced symmetry of the equations whereby one can scale $H$, $V$ and $\dot{\phi}_\infty$ to obtain new solutions of the system without the need 
to scale $B$.}. This requires a 
separate analysis and so we will first treat the general three-dimensional phase space dynamics and return to this later.

Next, one can express some useful quantities in terms of these variables:
\begin{align}\centering
\Omega_\phi&= \frac{\dot{\phi}_\infty^2}{6H^2}+ \frac{V(\phi)}{3H^2}=x^2+y^2\label{eq:omphi},\\
\omega_\phi &= \frac{\dot{\phi}^2_\infty-2V}{\dot{\phi}^2_\infty+2V}=\frac{x^2-y^2}{x^2+y^2}\label{eq:wphi},\\
D&\equiv\frac{2B^2X}{\Lambda^2}=6x^2z^2\quad\textrm{and}\\
Q&=\frac{\alpha +z^2 \left[\sqrt{6}x+6 x^2 (\gamma -\alpha )+3 \sqrt{\frac{3}{2}} x \left(y^2-x^2-1\right)\right]}{1-6x^2z^2}\label{eq:Qcosmo}.
\end{align}
$\Omega_\phi\sim1$ corresponds to a dark energy dominated solution, $Q=0$ corresponds to the absence of any fifth-forces on small scales and $D=1$ 
signifies a singularity in the Jordan frame metric (which is proportional to $(1-D)^{1/2}$). The problem of finding dark energy dominated 
accelerating 
solutions where fifth-forces are absent is then 
reduced to finding stable fixed points of the autonomous system where $\Omega_\phi$ and $\omega_\phi$ are compatible with current dark energy 
observations and $Q=0$. We treat solutions where $D=1$ as unviable for the reasons explained in section \ref{sec:via}.

\subsection{Fixed Points}

Before studying the new system (\ref{eq:eqx})--(\ref{eq:eqz}), it is worth recalling some useful properties of the equivalent system when only a 
conformal coupling is present, which corresponds to $z=0$. This was studied by \cite{Gumjudpai:2005ry} (see also \cite{Copeland:2006wr}). There are 
two stable fixed 
points:
\begin{itemize}
 \item {\bf Dark energy dominated fixed point}: This has $\Omega_\phi=1$ and $\omega_\phi=-1+\lambda^2/3$. It exists whenever $\lambda<\sqrt{6}$ and 
is stable whenever $\lambda<(\sqrt{\alpha^2+12}-\alpha)/2$.
\item {\bf Variable fixed point}: This may or may not give dark energy domination depending on the choice of parameters, although solutions which 
match the current observations suffer from the lack of a matter dominated era, at least when $\alpha$ and $\lambda$ are constant 
\cite{Amendola:1999er}. It is always stable 
but only exists when the dark energy dominated solution is unstable and is hence the only stable fixed point when $\lambda>\sqrt{6}$. There is a 
critical value of $\alpha$ below which it is a stable node and above which it is a stable spiral.
\end{itemize}

In the case of a purely conformal coupling, the phase space is a semi-circle in the $x$--$y$ plane (we do not consider $V(\phi)<0$). Since $z>0$ 
($B>0$ for stability reasons, see \cite{Zumalacarregui:2012us}) the phase-space of the disformal system is an infinite semi-circular prism restricted 
to the upper half of the $z$ plane ($z>0$). This parametrisation could be problematic because $z$ can potentially reach $\infty$ and so there may be 
fixed points at infinity. Indeed, this is the case but we will analyse the fixed points of the system 
(\ref{eq:eqx})--(\ref{eq:eqz}) first in order to make contact with the purely conformal case. Setting equations (\ref{eq:eqx})--(\ref{eq:eqz}) equal 
to zero, we find the fixed points given in table \ref{tab:zattracts}. The cosmological relevant parameters are given in table \ref{tab:zattracts2}. 
Points (1)--(5) are the fixed points found by \cite{Gumjudpai:2005ry} for the purely conformal case, they have $z=0$ so that all disformal effects 
are absent\footnote{Note that $z=0$ does not necessarily mean that $B=0$. Since $z\propto H$ these points may correspond to the infinite future 
where $H=0$.}. Note also that these points have $Q=\alpha$ and hence lead to large unscreened fifth-forces on all scales unless one tunes this to 
very small values. Since $D=0$ at these points the Jordan frame metric is non-singular at late times. 

The new points are (6) and (7), which are both unviable as cosmological solutions. These points behave as 
a stiff fluid ($\omega_\phi=1$) and so cannot describe a dark energy dominated universe. Furthermore, point (7) is unphysical except for the special 
point $\gamma=\sqrt{3/2}$; when $\gamma>\sqrt{3/2}$ one has $\Omega_\phi>1$ and the point lies outside the physical state space. Even if this were 
not the case, they have $D=1$ signalling a metric singularity in the infinite future and are unviable for the reasons discussed in section 
\ref{sec:via}.

\begin{table*}
 \centering
\begin{tabular}{|c|c|c|c|c|}\hline
Name & $x$ &$y$ &$z$ &Existence \\\hline
\hline
(1)& $-\sqrt{\frac{2}{3}}\alpha $ & $0 $ &$0$&$\alpha<\sqrt{3/2}$ \\\hline
(2)& $-1 $ & $0 $ & $0 $&All\\\hline
(3)& $1 $ & $0 $ & $0 $&All\\\hline
(4)& $\frac{\lambda}{\sqrt{6}} $ & $\sqrt{1-\frac{\lambda^2}{6}} $ & $0 $&$\lambda<\sqrt{6}$\\\hline
(5)& $\frac{\sqrt{\frac{3}{2}}}{\alpha +\lambda } $ & $\frac{\sqrt{3+2 \alpha  (\alpha +\lambda )}}{\sqrt{2} (\alpha +\lambda )} $
& $0 $&$\alpha>3/\lambda-\lambda$\\\hline
(6)& $\sqrt{\frac{2}{3}}\gamma-\sqrt{\frac{2}{3}\gamma^2-1} $ & $0 $ & $\frac{1}{3} \sqrt{2\gamma^2
+\gamma\sqrt{4 \gamma ^2-6}-\frac{3}{2}} $ & $\gamma>\sqrt{3/2}$ \\\hline
(7)& $\sqrt{\frac{2}{3}}\gamma+\sqrt{\frac{2}{3}\gamma^2-1} $ & $0 $ & $\frac{1}{3} \sqrt{2\gamma^2
-\gamma\sqrt{4 \gamma ^2-6}-\frac{3}{2}} $&$\gamma=\sqrt{3/2}$ \\\hline
\end{tabular}\caption{The fixed points of the system (\ref{eq:eqx})--(\ref{eq:eqz}) with $\gamma\ne\lambda/2$. }\label{tab:zattracts}
\end{table*}

 \begin{table*}
 \centering
\begin{tabular}{|c|c|c|c|c|}\hline
Name & $Q$ &$\Omega_\phi$ & $\omega_\phi$ & $D$ \\\hline
\hline
(1)&$\alpha$&$1$&$1$&$0$\\\hline
(2)&$\alpha$&$1$&$1$&$0$\\\hline
(3)&$\alpha$&$\frac{2\alpha^2}{3}$&$1$&$0$\\\hline
(4)& $\alpha$&$1$&$-1+\frac{\lambda^2}{3}$&$0$\\\hline
(5)& $\alpha$&$\frac{3+\alpha  (\alpha +\lambda )}{(\alpha +\lambda )^2}$&$-1+\frac{3}{3+\alpha  (\alpha +\lambda )}$&$0$\\\hline
(6)&$-$&$\frac{1}{3} \left(\sqrt{-3+2 \gamma ^2}-\sqrt{2} \gamma\right)^2$&$1$&$1$\\\hline
(7)& $-$&$\frac{1}{3} \left(\sqrt{-3+2 \gamma ^2}+\sqrt{2} \gamma\right)^2$&$1$&$1$\\\hline
\end{tabular}\caption{The cosmological quantities at the fixed points of the system (\ref{eq:eqx})--(\ref{eq:eqz}) with $\gamma\ne\lambda/2$. Models 
with $D=1$ do not have a corresponding value of $Q$ since it is not determined uniquely by the properties of the fixed point (see the discussion in 
section \ref{sec:via}.
}\label{tab:zattracts2}
\end{table*}
 
\begin{figure}[ht]\centering
\includegraphics[width=0.45\textwidth]{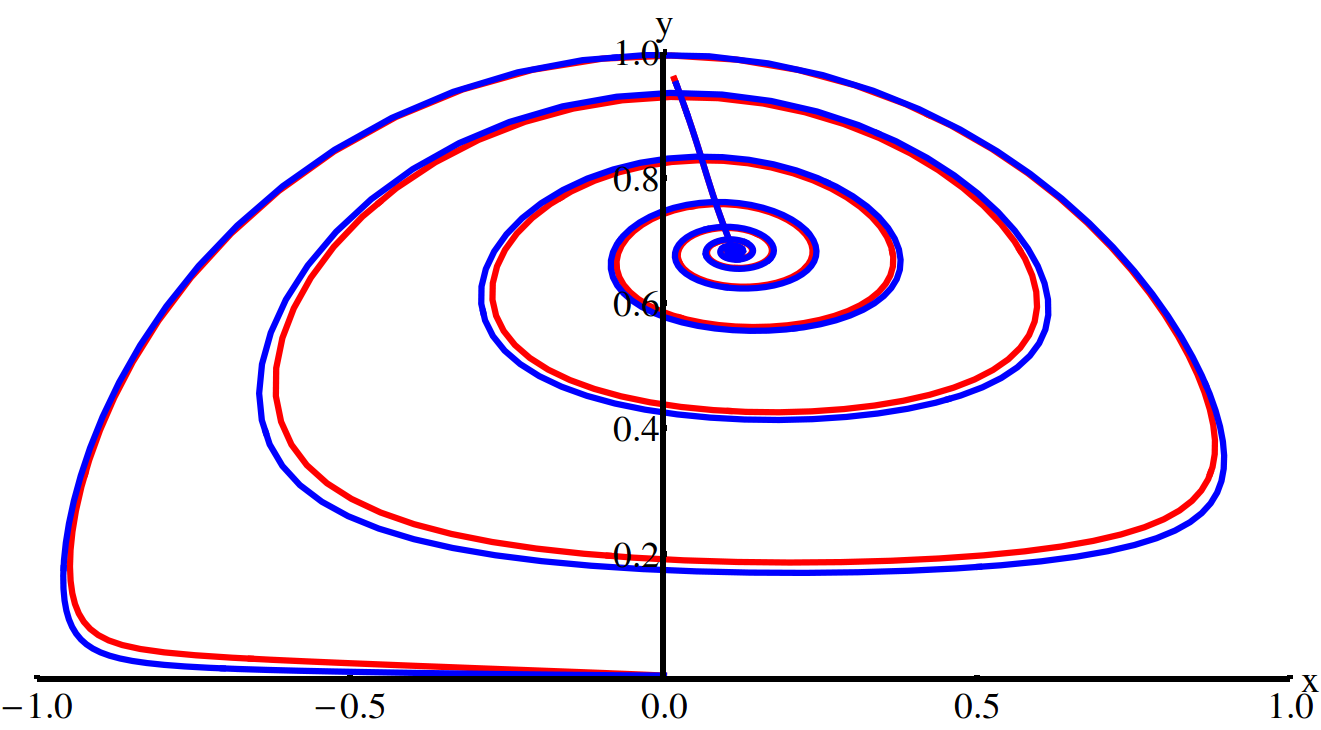}
\caption{The trajectories in the $x$--$y$ plane for $\alpha=5$, $ \gamma=4$ and $\lambda=6$ with initial conditions $\phi_0=-1$ 
and $\phi^\prime_0=0$. The red track corresponds to $H_0/\Lambda=10^{-2}$ and the blue track to $H_0/\Lambda=10^{-3}$.}\label{fig:genspir}
\end{figure}

\begin{figure}[ht]\centering
\includegraphics[width=0.45\textwidth]{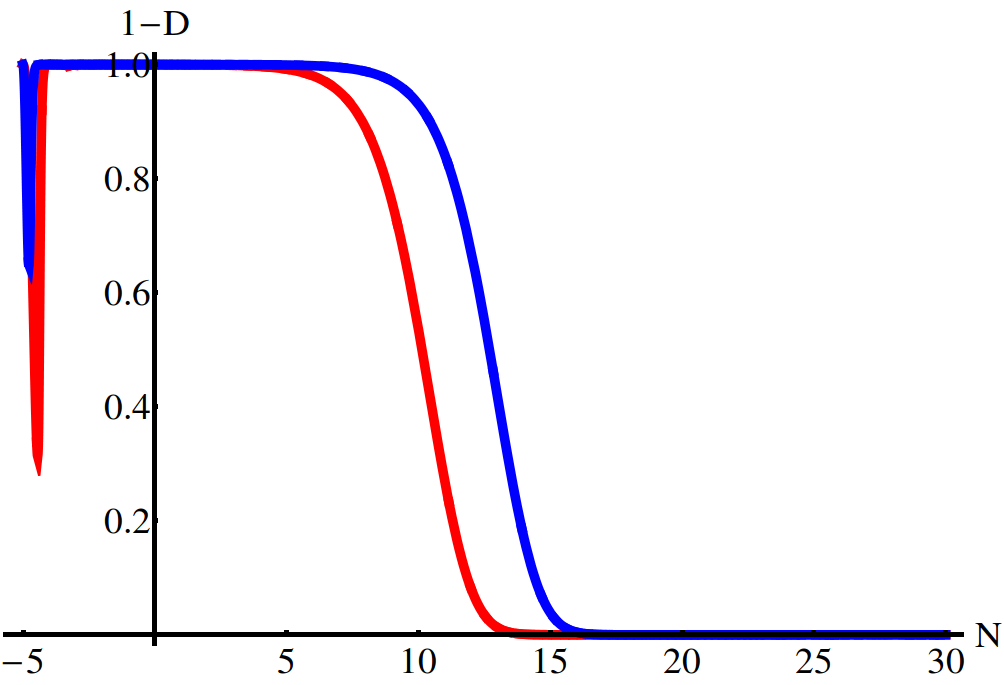}
\caption{$1-D$ as a function of $N$ for the models described in figure \ref{fig:genspir}. Recall that 
$\sqrt{-\tg}=(1-D)^{1/2}\sqrt{-g}$ and so $D=1$ represents a metric singularity of the Jordan frame metric.}\label{fig:detgen}
\end{figure}

Since the dimension of the phase space is increased from the purely conformal case, the stability of the fixed points is altered; there are now 
three eigenvalues instead of two. In particular, 
whereas the purely conformal fixed points ((1)--(5)) do not depend on $\gamma$, their stability does. The eigenvalues, $e_{1,2,3}$ for 
each fixed point are given in Appendix \ref{sec:ape}. One may check that there are choices of the parameters $\alpha$, 
$\gamma$ and $\lambda$ where none of these fixed points are stable. Indeed, one may check that when $\alpha=5$, $ \gamma=4$ and $\lambda=6$, all of 
the fixed points are either unstable or saddle points. The trajectory in the 
$x$--$y$ plane of solutions with $H_0/\Lambda=10^{-2}$ and $10^{-3}$ and initial conditions $\phi_0=-1$ and $\phi^\prime_0=0$ is shown in figure 
\ref{fig:genspir}. Furthermore, we show the evolution of $(1-D)$ in figure \ref{fig:detgen}. One can see that in each case, the system spends a long 
time near the saddle point before moving towards $x=0$, $y=1$. Along this trajectory, $(1-D)$ approaches zero so that the universe 
is tending to a state where the Jordan frame metric is singular. Furthermore, one can see that this behaviour is independent of the choice of 
$\Lambda$ and one can check that it is not affected by the initial conditions either. This suggests that the system has a common late-time behaviour 
not captured by the fixed-point analysis above. The next subsection is devoted to understanding this.
\subsection{Fixed Points at Infinity}

Numerically, one finds that a continued integration of the system to later times pushes $x$ closer to $0$, $y$ closer to $1$ and $z$ to increasingly 
larger values. This suggests that the behaviour may be due to fixed points at $z=\infty$ and so we investigate these by compactifying the phase 
space. Defining
\begin{equation}
 Z\equiv \frac{z}{z+1}
\end{equation}
so that $0\le Z\le 1$ we can map the points at $z=\infty$ to $Z=1$ whilst $z=0$ corresponds to $Z=0$. The phase space in these coordinates is 
then 
the finite semi-circular prism. In these coordinates, the system is described by the three first-order equations
\begin{widetext}
\begin{align}
 \frac{\dd x}{\dd N}&=-3 x+\sqrt{\frac{3}{2}} \lambda  y^2+\frac{3}{2} x \left(x^2-y^2+1\right)\nonumber\\&+\sqrt{\frac{3}{2}}\frac{ 
\left(x^2+y^2-1\right) \left(6 x^2 Z^2 (\alpha -\gamma )+3 \sqrt{6} x Z^2-3 \lambda  y^2 
Z^2-\alpha  (Z-1)^2\right)}{Z^2 \left(9 x^2+3 y^2-4\right)+2 Z-1},\label{eq:eqnxc}\\ 
\frac{\dd y}{\dd N}&=\frac{3}{2} y \left(1+x^2-y^2\right)-\sqrt{\frac{3}{2}} x y
 \lambda\label{eq:eqyc}\quad\textrm{and}\\ \frac{\dd Z}{\dd N}&=\frac{1}{2} (Z-1) Z \left(3 x^2-2 \sqrt{6} \gamma  x-3 y^2+3\right).\label{eq:eqzc}
\end{align}
\end{widetext}
In addition to the fixed points found above, one finds the fixed points given in table \ref{tab:comp1}\footnote{We have not included fixed points 
where $Z>1$ or 
$Z<0$, which lie outside the physical state space. Furthermore, one finds that there is an additional fixed point when $\gamma\rightarrow\infty$ 
where $Z=1$, which we also ignore. This may be relevant for models where $\gamma$ can reach infinity.}. The cosmological parameters at each point are 
shown in table \ref{tab:comp2}. We have omitted point (11) since this is discussed in detail in the next section.
\begin{table*}
 \centering
\begin{tabular}{|c|c|c|c|c|}\hline
Name & $x$ &$y$ &$Z$ &Existence \\\hline
\hline
(8)& $-1 $ & $0 $ &$1$&All \\\hline
(9)& $0 $ & $0 $ & $1 $&All\\\hline
(10)& $1 $ & $0 $ & $1 $&All\\\hline
(11)& $0 $ & $1$ & $1 $&All\\\hline
(12)& $-\frac{\sqrt{2 (\alpha -\gamma )^2-9}+\sqrt{2} \alpha -\sqrt{2} \gamma }{3 \sqrt{3}}$ & $0$
& $1 $&$\alpha-\gamma>\frac{3}{\sqrt{2}}$\\\hline
(13)& $\frac{\sqrt{2 (\alpha -\gamma )^2-9}-\sqrt{2} \alpha +\sqrt{2} \gamma }{3 \sqrt{3}} $ & $0 $ & $1 $ & $\alpha-\gamma>\frac{3}{\sqrt{2}}$ 
\\\hline
(14)& $\frac{\lambda}{\sqrt{6}}$ & $\sqrt{1-\frac{\lambda^2}{6}} $ & $1 $&$\lambda<\sqrt{6}$ \\\hline
(15)& $\frac{\sqrt{6}}{2 \alpha -2 \gamma +3 \lambda }$ & $\frac{\sqrt{\frac{6}{2 \alpha -2 \gamma +3 \lambda }+2 \alpha -2 \gamma +\lambda 
}}{\sqrt{2 
\alpha -2 \gamma +3 \lambda }}  $ & $1 $&$2 \alpha -2 \gamma +3 \lambda >0$ \\\hline
(16)& $\frac{\sqrt{4 \gamma ^2-6}+2 \gamma }{\sqrt{6}}$ & $0$ & $\frac{1}{\sqrt{4 \gamma ^2-6}+2 \gamma +1} $&$\gamma>\sqrt{\frac{2}{3}}$ \\\hline
(17)& $\frac{2 \gamma -\sqrt{4 \gamma ^2-6}}{\sqrt{6}}$ & $0 $ & $\frac{\sqrt{4 \gamma ^2-6}+2 \gamma +1}{4 \gamma +7} $&$\gamma>\sqrt{\frac{2}{3}}$ 
\\\hline
\end{tabular}\caption{The fixed points of the compactified system system (\ref{eq:eqnxc})--(\ref{eq:eqzc}). Fixed points with $Z=0$ are not shown 
since they are already present in table \ref{tab:zattracts}. Similarly, fixed points (4) and (5) are not shown. We have listed the condition for the 
existence of the fixed point but that is not to say that it lies inside the physical phase space; one should check with table \ref{tab:comp2} to 
ensure that the cosmological parameters, especially $\Omega_\phi$, assume values inside the physical state space. }\label{tab:comp1}
\end{table*}
\begin{table*}
\begin{tabular}{|c|c|c|c|c|}\hline
Name & $Q$ &$\Omega_\phi$ & $\omega_\phi$ & $D$ \\\hline
\hline
(8)&$\pm\infty$&$1$&$1$&$\infty$\\\hline
(9)&$\alpha$&$0$&$0$&$\infty$\\\hline
(10)&$\pm\infty$&$1$&$1$&$\infty$\\\hline
(12)&$\pm\infty$&$\frac{1}{27} \left(\sqrt{2 (\alpha -\gamma )^2-9}+\sqrt{2} \alpha -\sqrt{2} \gamma \right)^2$&$1$&$\infty$\\\hline
(13)&$\pm\infty$&$\frac{1}{27} \left(\sqrt{2 (\alpha -\gamma )^2-9}+\sqrt{2} \alpha -\sqrt{2} \gamma \right)^2$&$1$&$\infty$\\\hline
(14)&$\pm\infty$&$1$&$-1+\lambda^2/3$&$\infty$\\\hline
(15)&$\pm\infty$&$\frac{4 \alpha ^2-8 \alpha  \gamma +8 \alpha  \lambda +4 \gamma ^2-8 \gamma  \lambda +3 \lambda ^2+12}{(2 \alpha -2 \gamma +3 
\lambda )^2}$&$-\frac{(2 \alpha -2 \gamma +\lambda ) (2 \alpha -2 \gamma +3 \lambda )}{4 \alpha ^2-8 \alpha  \gamma +8 \alpha  \lambda +4 \gamma ^2-8 
\gamma  \lambda +3 \lambda ^2+12}$&$\infty$\\\hline
(16)&$-$&$\frac{1}{6} \left(\sqrt{4 \gamma ^2-6}+2 \gamma \right)^2$&$1$&$1$\\\hline
(17)&$-$&$\frac{1}{6} \left(\sqrt{4 \gamma ^2-6}-2 \gamma \right)^2$&$1$&$1$\\\hline
\end{tabular}\caption{The cosmological quantities at the fixed points of the compactified system. Note that $Q\rightarrow\pm\infty$ 
reflects the fact that $z\rightarrow\infty$. The sign depends on the choice of $\gamma$, $\lambda$ and $\alpha$. Models 
with $D=1$ do not have a corresponding value of $Q$ since it is not determined uniquely by the properties of the fixed point (see the discussion in 
section \ref{sec:via}}\label{tab:comp2}
\end{table*}
The eigenvalues are listed in Appendix \ref{sec:appcm} but here we note that the eigenvalues for point (11) are
\begin{equation}
 e_1=-3,\quad e_2=0\quad\textrm{and}\quad e_3=0.
\end{equation}
%
%

\subsubsection{Centre Manifold Analysis}

\begin{figure}[ht]\centering
\includegraphics[width=0.45\textwidth]{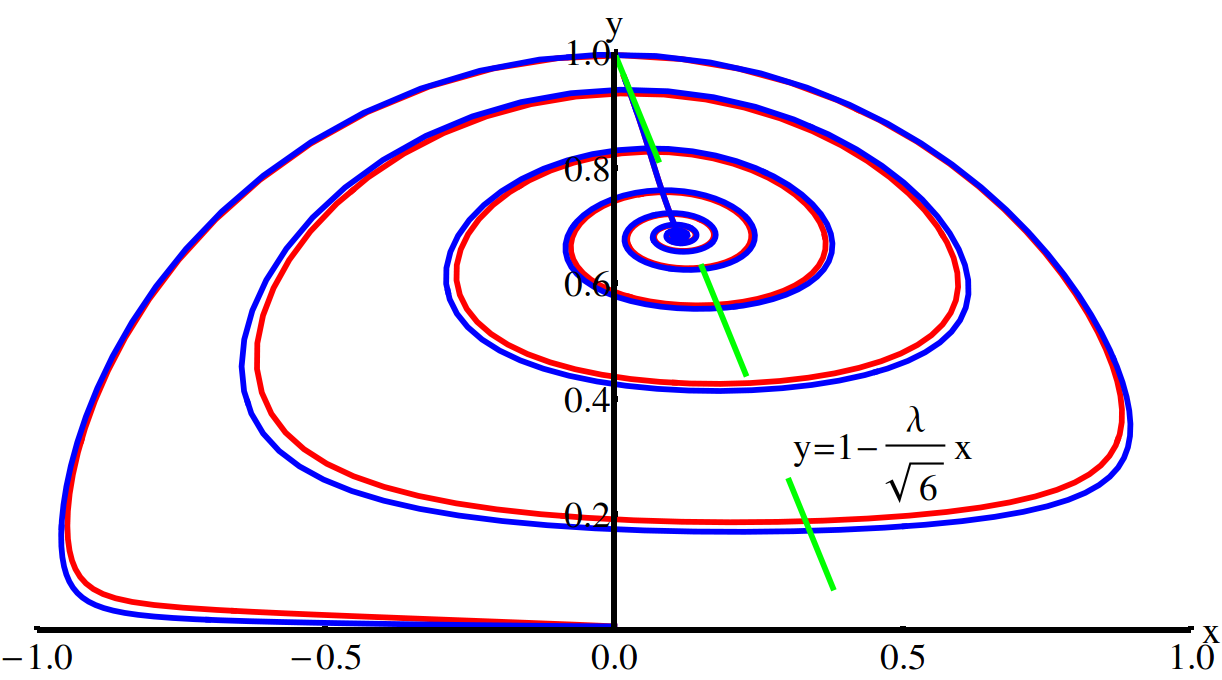}
\caption{The $x$--$y$ plane for the same models as figure \ref{fig:genspir} but overlaid with the centre manifold prediction for 
the late-time trajectories $y=1-\lambda x/\sqrt{6}$ (green dashed line).}\label{fig:spir2}
\end{figure}

The two zero eigenvalues indicate that a linear analysis is not sufficient to determine the stability of the system 
near this fixed point. In order to do this, one must perform a centre manifold analysis. This has been used in cosmological systems previously (see 
\cite{Alho:2014fha} for a recent application to $m^2\phi^2$ potentials and references therein for further examples) and we give a full account for 
the unfamiliar reader in Appendix \ref{sec:appcm}. For simplicity, we move the point to the origin by making the change of variables:
\begin{align}
 Z&=W+1\label{eq:W}\quad\textrm{and}\\
 y&=Y+1.\label{eq:Y}
\end{align}
We will not give the new first-order system here but for completeness it is given in Appendix \ref{sec:xYW}. This change of variables does not change 
the eigenvalues but does change the eigenvectors, which, in this basis, are
\begin{equation}
 \vec{e}_1=\colvec[0]{1}{0},\quad\vec{e}_2=\colvec[0]{0}{1}\quad\textrm{and}\quad\vec{e}_3=\colvec[-\frac{\sqrt{6}}{\lambda }]{1}{0},
\end{equation}
where the first is an eigenvalue $-3$ vector and the final two are zero-eigenvectors. As discussed in Appendix \ref{sec:dyna}, perturbations about 
the 
fixed point will evolve according to:
\begin{equation}
 \colvec[\delta x]{\delta Y}{\delta W}=\vec{B}_1e^{e_1t}+\vec{B}_2e^{e_2t}+\vec{B}_3e^{e_3t},
\end{equation}
where $\vec{B}_i$ are the eigenbasis of the matrix $M_{ij}$ defined in (\ref{eq:matdef}), which is found by expanding the equations to linear order 
in 
small perturbations $\{\delta x, \delta Y, \delta W\}$. One can see that zero eigenvalues neither grow nor decay at linear order, one must go to the 
next-to-leading order to determine the dynamics. This means that for any initial configuration sufficiently close to the fixed point, the trajectory 
along the $\vec{e}_1$ direction will rapidly tend to its 
fixed point value whereas the trajectory in the plane described by $\vec{e}_{1,2}$---the centre manifold---will evolve on a slower time-scale. The 
essence 
of the centre manifold technique is to find the new variable along the $\vec{e}_1$ direction and assume that its evolution equation is already 
minimised. One 
can then solve this to obtain an algebraic expression for this variable in terms of those parametrising the centre manifold. The equations for the 
other two variables---the \textit{centre variables}---then describe the dynamics in the centre manifold at late times. One has then reduced the 
dimension of the phase space to that of the centre manifold. Writing
\begin{equation}
 \colvec[x]{Y}{W}=\sum_i A_i \vec{e}_i,
\end{equation}
one finds
\begin{equation}
 \colvec[x]{Y}{W}=\colvec[-\frac{\sqrt{6}}{\lambda}A_3]{A_1+A_3}{A_2},
\end{equation}
which can be inverted to give
\begin{equation}
 \colvec[A_1]{A_2}{A_3}=\colvec[Y+\frac{\lambda}{\sqrt{6}}x]{W}{-\frac{\lambda}{\sqrt{6}}x}.
\end{equation}
Setting $A_1^\prime=0$ i.e. the variable along $\vec{e}_1$, one finds
\begin{equation}
 \frac{\dd y}{\dd x}=-\frac{\lambda}{\sqrt{6}}.
\end{equation}
Therefore, at late times, when the system evolves towards fixed point (11) the trajectory in the $x$--$y$ plane is
\begin{equation}
 y\approx1-\frac{\lambda}{\sqrt{6}}x.
\end{equation}
In figure \ref{fig:spir2} we plot the same models as shown in figure \ref{fig:genspir} and overlay this line. One can see that the trajectories 
indeed converge to this at late times once they have left the saddle point. 

Finally, one can examine the behaviour of the determinant of the Jordan frame metric along this trajectory. We begin with the equations for the 
centre variables, where we have set $A_1=0$, which follows from the values of $x$ and $Y$ at the fixed point:
\begin{widetext}
\begin{align}
 \frac{\dd A_2}{\dd N}&=-\frac{3 A_2 (A_2+1) A_3 \left(A_3 \left(\lambda ^2-6\right)+2 \lambda  (\lambda -2 \gamma )\right)}{2 
\lambda ^2}\label{eq:A2}\quad\textrm{and}\\
\frac{\dd A_3}{\dd N}&=\nonumber-\frac{\lambda A_3 \left(\frac{6 A_3}{\lambda ^2}+A_3+2\right) \left((A_2+1) \left(2 \alpha 
+(A_2+1) \left(-\alpha +\frac{18 A_3 (2 A_3 (\alpha -\gamma )-\lambda )}{\lambda ^2}-3 (A_3+1)^2 \lambda \right)\right)-\alpha 
\right)}{2(A_2+1) \left((A_2+1) \left(A_3^2 \left(\frac{54}{\lambda ^2}+3\right)+6 A_3-1\right)+2\right)-2}\\\label{eq:A3}&-\frac{3 A_3^2 
\left(A_3 \left(\lambda ^2-6\right)+2 \lambda ^2\right)}{2\lambda ^2}-3 A_3-\frac{(A_3+1)^2 \lambda^2}{2}.
\end{align}
\end{widetext}
One can then find the fixed points of this reduced phase space. The only two physical fixed points are
\begin{equation}
 A_2=\frac{12 (\lambda -2 \gamma )}{24 \gamma \pm\lambda ^2-12 \lambda \pm6}\quad\textrm{and}\quad A_3=\frac{2 \lambda  (2 \gamma -\lambda )}{\lambda 
^2-6}.
\end{equation}
In these coordinates, one has 
\begin{equation}
 D=1-\frac{36A_3^2(A_2+1)^2}{\lambda^2A_2^2}
\end{equation}
and one can verify that this is indeed zero at these fixed points. Therefore, any trajectory in the $x$--$y$ plane that evolves towards this fixed 
point will evolve in such a way that $\sqrt{-\tg}\rightarrow0$ but only in the infinite future. This is the underlying reason for the pathology 
resistance discussed above\footnote{One should note that we refer to this phenomena as \textit{pathology resistance} in order to make contact with 
previous works. We consider models exhibiting metric singularities to be unviable for the reasons discussed in section \ref{sec:via}. One should 
note that it is possible to construct theories where the visible sector coupled to the Einstein frame metric but dark matter couples to the Jordan 
frame metric. In this case, it is less clear whether these solutions are indeed pathological since the non-relativistic limit is well-defined for 
visible matter and fifth-forces are not an issue.}: the system will evolve towards a metric singularity but only in the infinite future and therefore 
any trajectory must 
necessarily slow down as it is approached. To date, this feature has been observed numerically but no underlying reason has been discerned. What we 
have shown here is that this is a generic feature of models where the conformal fixed points are all unstable. As discussed in section \ref{sec:via}, 
models that exhibit this singularity are not viable since they suffer from large fifth-forces and may not have a viable non-relativistic limit and so 
we conclude that in the general case, the only viable models are the disformal generalisations of the two conformal fixed points ((4) and (5)). 
Therefore, one requires either $\alpha\ll1$ or that the visible sector is coupled to the Einstein frame metric directly in order to avoid 
fifth-forces. Furthermore, the late-time properties of these models are identical to those of the equivalent purely conformal theory, although the 
dynamics before the fixed point is reached may be different. 


\section{Reduced Phase Space}\label{sec:red}

In this section, we examine the special parameter choice $\gamma=\lambda/2$, which we argued above reduces the dimension of the phase 
space to two and therefore requires a separate analysis. We can choose to eliminate either $z$ or $y$ from the equations and we choose to eliminate 
$z$ in order to make contact with the purely conformal case. Furthermore, $z\rightarrow\infty$ corresponds to $y\rightarrow0$ and so this 
substitution captures all of the new fixed points without having to compactify further. Using equation (\ref{eq:ps2}) in equations (\ref{eq:eqx}) and 
(\ref{eq:eqy}) we find the 
following two-dimensional autonomous system for $x$ and $y$:
\begin{widetext}
 \begin{align}
  \frac{\dd x}{\dd N}=\sqrt{\frac{3}{2}}\frac{\left(x^2+y^2-1\right) \left(x^2 (2 \alpha -\lambda )+\sqrt{6} x-y^2 \left(\alpha  \lt 
^2+\lambda \right)\right)}{3 x^2-\left(\lt ^2-1\right) y^2-1}+\frac{3}{2} x \left(x^2-y^2+1\right)-3 x+\sqrt{\frac{3}{2}} \lambda  y^2,
 \end{align}
\end{widetext}
where the second equation is (\ref{eq:eqy}) and $\lt\equiv\Lambda/m_0$. The fixed points which lie inside the physical phase space are shown in table 
\ref{tab:specs} and the 
cosmological quantities at these points are shown in table \ref{tab:spec2}. Note that one has a choice of minimising either the $x$--$y$ or $x$--$z$ 
system. Either is fine, but it is important to note that the redundant equation acts as a constraint and should be identically zero at the fixed 
points. When this is not the case the points are unphysical and we do not include them. Points (18) and (19) are points (4) and (5), although they 
have different 
scalar charges and their stability is different; the interesting point is (20). We list the eigenvalues in Appendix \ref{sec:ape}. The stability of 
point (18) is identical to the conformal case but the stability of (19) is altered from both the conformal case and the 
case where $\gamma\ne\lambda/2$. Furthermore, one can check that point (20) is stable over a large range of parameter space and is often 
simultaneously stable when point (18) is. When point (18) does not exist, it is often the case that point (19) is unstable and (20) is the only 
stable attractor of the system. 
\begin{table*}
 \centering
\begin{tabular}{|c|c|c|c|}\hline
Name & $x$ &$y$ &Existence \\\hline
\hline
(18)&$\frac{\lambda}{\sqrt{6}} $ & $\sqrt{1-\frac{\lambda^2}{6}} $ & $\lambda<\sqrt{6}$
\\\hline
(19)& $\frac{\sqrt{\frac{3}{2}}}{\alpha +\lambda } $ & $\frac{\sqrt{3+2 \alpha  (\alpha +\lambda )}}{\sqrt{2} (\alpha +\lambda )} $
& $\alpha>3/\lambda-\lambda$\\\hline
(20)& $\frac{\lambda  \lt ^2-\sqrt{\left(\lambda ^2-6\right) \lt ^4+12 \lt ^2}}{\sqrt{6} \left(\lt ^2-2\right)}$&$\sqrt{6} 
\left({\left(\lambda ^2-3\right) \lt ^2+\lambda  \sqrt{\lt ^2 \left(\left(\lambda ^2-6\right) \lt 
^2+12\right)}+6}\right)^{-1/2}$&$\lambda>\sqrt{3}$\\\hline
\end{tabular}\caption{The fixed points of the system when $\gamma=\lambda/2$. Note that the existence indicates where the point 
exists; one may find parameter choices where the point exists but lies outside the physical state space. Note that point (20) exists 
for all $\lt$ when $\lambda>\sqrt{3}$. When the converse is true one can find a narrow region in the $\lt$--$\lambda$ plane 
where the point still exists. We will not be interested in this region in this work.}\label{tab:specs}
\end{table*}
\begin{table*}
\begin{tabular}{|c|c|c|c|c|}\hline
Name & $Q$ &$\Omega_\phi$ & $\omega_\phi$ & $D$ \\\hline
\hline
(18)& $\frac{\alpha  \left(\lambda ^2 \left(\lt ^2+2\right)-6 \lt ^2\right)}{\left(\lambda ^2-6\right) \lt 
^2}$&$1$&$-1+\frac{\lambda^2}{3}$&$-\frac{2 \lambda ^2}{\left(\lambda ^2-6\right) \lt ^2}$\\\hline
(19)& $\alpha  -\frac{6}{\lt ^2 \left(2 \alpha ^2+2 \alpha  \lambda 
+3\right)}$&$\frac{\alpha  (\alpha +\lambda )+3}{(\alpha +\lambda )^2}$&$-1+\frac{3}{3+\alpha  (\alpha +\lambda )}$&$\frac{6}{\lt ^2 (2 \alpha  
(\alpha +\lambda )+3)}$\\\hline
(20)&$-$ &$\frac{3 \left(\lt ^2+2\right)}{\left(\lambda ^2-3\right) \lt ^2+\lambda  \sqrt{\left(\lambda ^2-6\right) \lt ^4+12 \lt 
^2}+6}$ &$1-\frac{4}{\lt ^2+2}$ &$1$\\\hline
\end{tabular}\caption{The cosmological quantities at the fixed points of the system when $\gamma=\lambda/2$. Models 
with $D=1$ do not have a corresponding value of $Q$ since it is not determined uniquely by the properties of the fixed point (see the discussion in 
section \ref{sec:via}}\label{tab:spec2}
\end{table*}

\begin{figure}[ht]\centering
\includegraphics[width=0.45\textwidth]{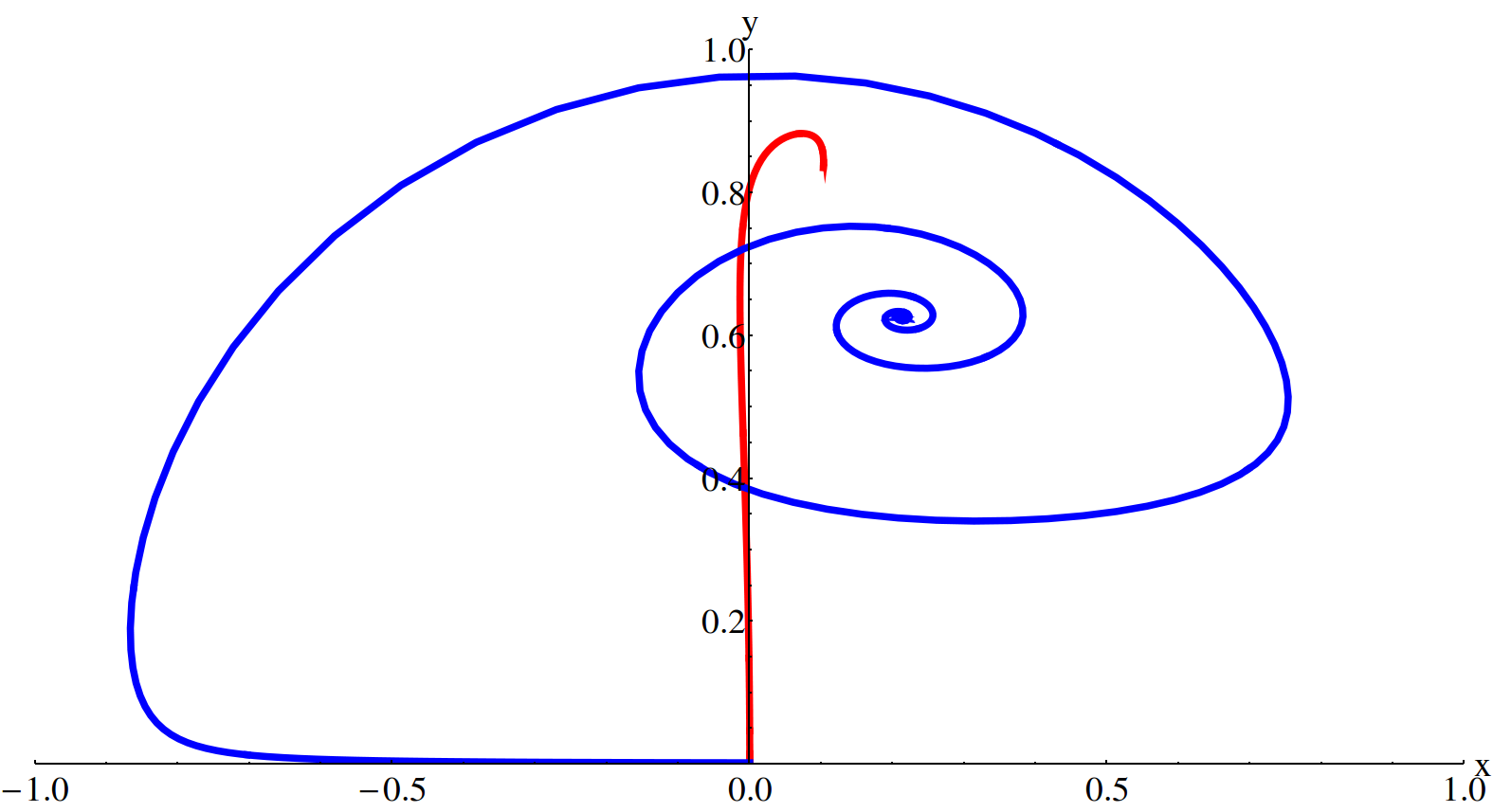}
\caption{The trajectories in the $x$--$y$ plane for a purely conformal theory with $\alpha=2$ (blue) and a disformal theory with 
$\gamma=\lambda/2$ (red). In each case the initial conditions are $\phi_0=-1$ and $\phi^\prime_0=0$. The parameters $\Lambda$, $m_0$ 
and $\lambda$ were chosen such that the disformal attractor corresponds to a universe where $\omega_\phi$ and $\Omega_\phi$ match 
the WMAP9 observations.}\label{fig:specatt}
\end{figure}

\begin{figure}[ht]\centering
\includegraphics[width=0.45\textwidth]{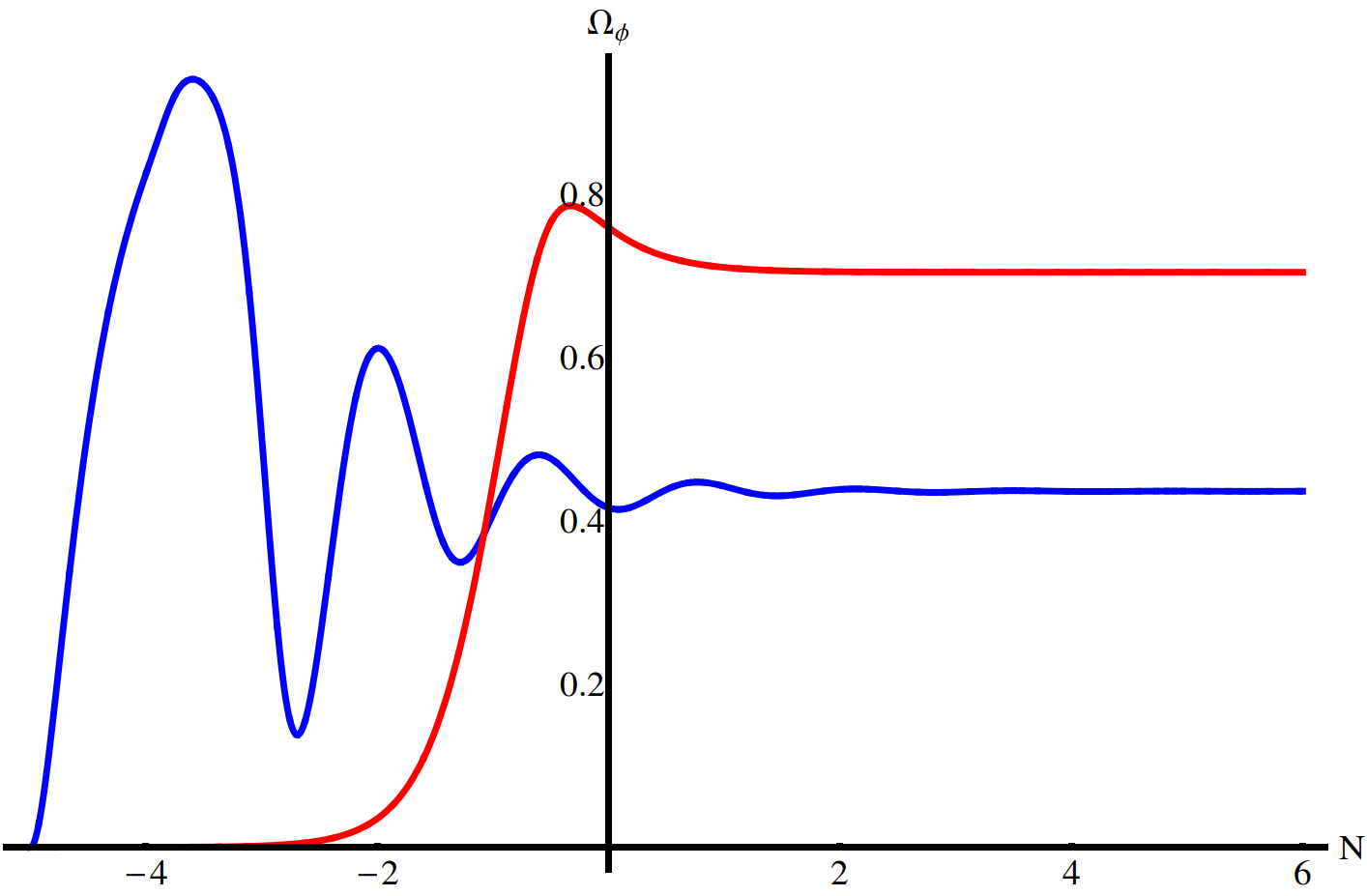}
\caption{$\Omega_\phi$ as a function of $N$ for a purely conformal theory with $\alpha=2$ (blue) and a disformal theory with 
$\gamma=\lambda/2$ (red). The parameters and initial conditions are as indicated in figure \ref{fig:specatt}'s caption.}\label{fig:specOm}
\end{figure}

\begin{figure}[ht]\centering
\includegraphics[width=0.45\textwidth]{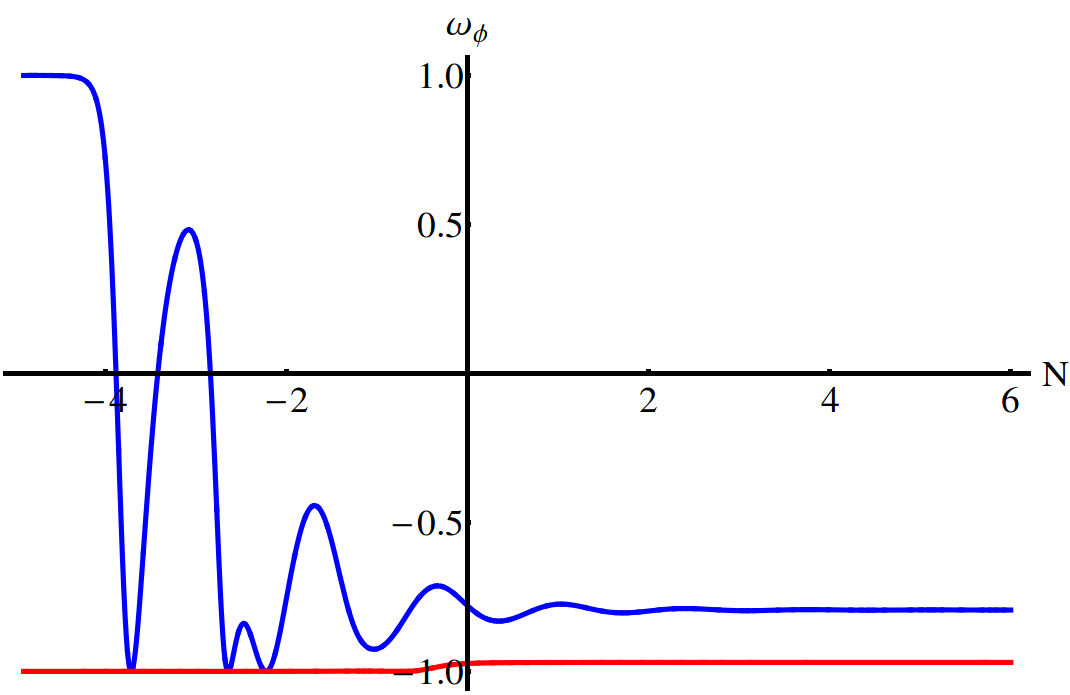}
\caption{$\omega_\phi$ as a function of $N$ for a purely conformal theory with $\alpha=2$ (blue) and a disformal theory with 
$\gamma=\lambda/2$ (red). The parameters and initial conditions are as indicated in figure \ref{fig:specatt}'s caption.}\label{fig:specw}
\end{figure}

Interestingly, this fixed point can match the current observations of dark energy.
To see this, consider as an example the WMAP9 results \cite{Hinshaw:2012aka}, $\omega=-0.97$, $\Omega_{\rm 
DE}=0.704$\footnote{Note that this data set assumes that $\omega$ is fixed whereas it varies in our model and so a more realistic method would 
to use a varying $\omega$ fit such as the $\omega_0$--$\omega_a$ parametrisation. The analysis here is a proof of principle only and so we will not 
concern ourselves with a more realistic data analysis.}. One finds that fixed point (20) can reproduce this exactly by taking $\lambda=3.77953$ and 
$\lt=0.174519$ with $\alpha$ arbitrary. As an example, consider the case $\alpha=2$. In the purely conformal case the system would evolve to the 
variable fixed point where $\Omega_\phi\approx0.436$ and $\omega_\phi\approx-0.794$, which are far from the WMAP values. In figure \ref{fig:specatt} 
we plot the $x$--$y$ plane for both the purely conformal case and the disformal case with the same initial conditions. One can see that the disformal 
trajectory converges to fixed point (20) whereas the purely conformal case reaches fixed point (18) (or rather, the conformal equivalent). To 
illustrate this further, we plot $\Omega_\phi$ and $\omega_\phi$ in figures \ref{fig:specOm} and \ref{fig:specw}. It is interesting to note that, 
unlike the purely conformal case where there is a lack of a matter dominated era \cite{Amendola:1999er}, the disformal system delays the onset of the 
field rolling and allows for a period of matter domination. This behaviour was also noted in \cite{Sakstein:2014isa}. 
Finally, one can see that the fixed point has $D=1$, and so there is a singularity of $\sqrt{-\tg}$ in the infinite 
future. This fixed point is then unviable and we are left with points (18) and (19), the disformal equivalent of points (4) and (5). Even though the 
scalar charge differs from $\alpha$, one still requires $\alpha\ll1$ for these points to be viable. If one tunes the model parameters such 
that $\tilde{\Lambda}^2=2\lambda^2/(6-\lambda^2)$ so that the scalar charge at point (18) is zero then one finds $D=1$ so that the metric singularity 
is present and the model is unviable. One is then left with tuning $\alpha\ll1$ or decoupling the scalar from visible matter. Tuning the parameters 
so that the scalar charge at point (19) vanishes sets $D=\alpha$ and we must again tune $\alpha\ll1$ in order to avoid the pathologies associated 
with the metric singularity.


\section{Discussion and Conclusions}\label{sec:concs}

In this work we have studied the solution space of disformal theories of gravity using dynamical systems techniques. Our ultimate goal was to find 
stable dark energy dominated solutions where the local scalar charge is zero at late times. The Jordan frame metric can become singular, indicating a 
potential instability of the theory. Several previous works have noted a \textit{natural resistance to pathology} where numerical solutions show that 
the field slows down as the singularity is approached. Here, we were able to explain this by showing 
that any singularity is only reached in an infinite amount of time. Despite this, we argued that models that exhibit this behaviour are still 
unviable (at least when the scalar couples to the visible sector) due to large unscreened fifth-forces and that such a singularity may indicate the 
lack of an appropriate non-relativistic limit.

We showed that, in the general case, the phase space of the system is three-dimensional rather than the two-dimensional phase space of the equivalent 
quintessence and purely conformal theories. It was found that there are no new viable fixed points of the system and furthermore there are 
parameter choices where the attractors that are reached in the purely conformal case are now saddle points. This is due to the increased dimension of 
the phase space, which changes the characteristic polynomial for the eigenvalues to a cubic equation rather than a quadratic, yielding a third 
solution that can be positive. Numerically, we observed that when the parameters assume these values, the trajectories spend a long time near the 
saddle points before evolving towards a dark energy dominated fixed point. The Jordan frame becomes singular along the approach to this fixed point 
and hence these models are not viable. Two of the eigenvalues at this fixed 
point are zero and so it was necessary to use centre manifold techniques to discern the late-time behaviour.

We identified a tuning in parameter space, $\gamma=\lambda/2$, where the dimension of the phase space is reduced to two and therefore the dynamics 
are altered. There is a fixed point---point (20)---of the reduced phase space not present in the purely conformal system. This point is a 
stable attractor and can match the presently observed dark energy parameters by tuning $\lambda$ and $\tilde{\Lambda}=\Lambda/m_0$. In particular, by 
taking $\lambda=3.77953$ and $\lt=0.174519$ we were able to reproduce the WMAP9 density parameter and equation of state. Unfortunately, this fixed 
point is not viable since the Jordan frame metric becomes singular in the infinite future.

We therefore conclude that the only viable models of disformal dark energy have late-time properties 
to the equivalent models where the disformal coupling is absent. Furthermore, since these models are not stable over the entire parameter 
space, the inclusion of a disformal factor greatly reduces the choice of allowed model parameters. When these models are not stable, the cosmological 
solutions evolve towards fixed points where the Jordan frame metric is singular and include all the pathologies associated with this. The inclusion 
of a disformal factor seems to have little to say about dark energy and its inclusion introduces further problems that require reducing the viable 
parameter space to solve. With this in mind, it is worth raising the question of whether it is worthwhile to pursue these theories as dark energy 
candidates further.

In this work we 
have only considered theories where all matter couples to the Jordan frame metric universally but this does not necessarily have to be the case and, 
as we have remarked above several times already, one could alleviate some of the problems associated with the metric singularity by coupling visible 
matter to the Einstein frame metric. In this case, the issue of fifth-forces is moot since objects in the solar system do not feel the fifth-force 
and there is no ambiguity as to how the coordinate time is related to the clock used by static observers. Furthermore, since the evolution of the 
universe 
in the matter era is driven by dark matter and not baryons, some of the fixed points considered unviable here may be viable in these more general 
theories. Of particular interest is fixed point (20), which can match the presently observed dark energy density and equation of state by tuning only 
two of the four free parameters (three if one counts the $\gamma=\lambda/2$ tuning). Whether or not this scenario is viable requires further 
investigation beyond the scope of this work. Dark matter particles will see an effective value of Newton's constant given by $(1+2Q^2)G$, which 
becomes infinite as the fixed point is approached and so one expects to see novel effects in the cold dark matter power spectrum and collapsed 
haloes. Another potential issue is whether the apparent lack of a non-relativistic limit is incompatible with the notion of non-relativistic cold 
dark 
matter altogether. 
%

Here we have taken the 
simplest model where $\gamma$, $\lambda$ and $\alpha$ are constants. The dynamics are altered when one allows them to vary. One can then investigate 
the dynamics of the new system, which may have a dimension larger 
than three, or try to find models where the late-time fixed points give the observed dark energy parameters. In terms of our model, the choice 
$\gamma=\lambda/2$ was a parameter tuning but for more general models one requires that $\gamma(\phi)=\lambda(\phi)/2$ is reached dynamically. It 
remains to be seen if one can contrive more general models where this is achieved. One may also worry that quantum corrections can spoil this 
tuning; this certainly merits investigation.

Only recently have the properties of disformal theories been fully elucidated and, unlike more well-studied models, there is no canonical disformal 
paradigm. Indeed, to date, a viable model that screens fifth-forces and can account for all cosmological observations is still lacking. What we have 
done here is to classify the possible cosmological solutions and identify a class of models---albeit using a special tuning---that can reproduce the 
observed dark energy parameters today. It is unlikely that this solution is viable when the scalar couples to visible matter and so the study of 
models where the scalar couples to dark matter only is certainly well-motivated. Future studies must assess how well these models can fit 
probes of the expansion history such as the supernova luminosity distance as well as linear probes such as the CMB. It would also be interesting to 
study the cosmology of more realistic models that include components neglected here such as radiation.

\section*{Acknowledgements}
I am incredibly grateful to Claes Uggla for several enlightening conversations and for bringing new techniques such as the centre manifold to my 
attention. The final version of this work benefited greatly from discussions with Ruth Gregory and Kazuya Koyama. I would also like to thank Tomi 
Koivisto and Antonio Padilla for some interesting discussions. Finally, I am grateful to Carsten van de Bruck and Nelson Nunes for pointing out some 
typographical errors in the final version. Research at Perimeter Institute is 
supported by the Government of Canada through Industry Canada and by the Province of Ontario through the 
Ministry of Economic Development \& Innovation.

\bibliography{ref}

\begin{thebibliography}{62}%
\makeatletter
\providecommand \@ifxundefined [1]{%
 \@ifx{#1\undefined}
}%
\providecommand \@ifnum [1]{%
 \ifnum #1\expandafter \@firstoftwo
 \else \expandafter \@secondoftwo
 \fi
}%
\providecommand \@ifx [1]{%
 \ifx #1\expandafter \@firstoftwo
 \else \expandafter \@secondoftwo
 \fi
}%
\providecommand \natexlab [1]{#1}%
\providecommand \enquote  [1]{``#1''}%
\providecommand \bibnamefont  [1]{#1}%
\providecommand \bibfnamefont [1]{#1}%
\providecommand \citenamefont [1]{#1}%
\providecommand \href@noop [0]{\@secondoftwo}%
\providecommand \href [0]{\begingroup \@sanitize@url \@href}%
\providecommand \@href[1]{\@@startlink{#1}\@@href}%
\providecommand \@@href[1]{\endgroup#1\@@endlink}%
\providecommand \@sanitize@url [0]{\catcode `\\12\catcode `\$12\catcode
  `\&12\catcode `\#12\catcode `\^12\catcode `\_12\catcode `\%12\relax}%
\providecommand \@@startlink[1]{}%
\providecommand \@@endlink[0]{}%
\providecommand \url  [0]{\begingroup\@sanitize@url \@url }%
\providecommand \@url [1]{\endgroup\@href {#1}{\urlprefix }}%
\providecommand \urlprefix  [0]{URL }%
\providecommand \Eprint [0]{\href }%
\providecommand \doibase [0]{http://dx.doi.org/}%
\providecommand \selectlanguage [0]{\@gobble}%
\providecommand \bibinfo  [0]{\@secondoftwo}%
\providecommand \bibfield  [0]{\@secondoftwo}%
\providecommand \translation [1]{[#1]}%
\providecommand \BibitemOpen [0]{}%
\providecommand \bibitemStop [0]{}%
\providecommand \bibitemNoStop [0]{.\EOS\space}%
\providecommand \EOS [0]{\spacefactor3000\relax}%
\providecommand \BibitemShut  [1]{\csname bibitem#1\endcsname}%
\let\auto@bib@innerbib\@empty
\bibitem [{\citenamefont {Riess}\ \emph {et~al.}(1998)\citenamefont {Riess}
  \emph {et~al.}}]{Riess:1998cb}%
  \BibitemOpen
  \bibfield  {author} {\bibinfo {author} {\bibfnamefont {A.~G.}\ \bibnamefont
  {Riess}} \emph {et~al.} (\bibinfo {collaboration} {Supernova Search Team}),\
  }\href {\doibase 10.1086/300499} {\bibfield  {journal} {\bibinfo  {journal}
  {Astron.J.}\ }\textbf {\bibinfo {volume} {116}},\ \bibinfo {pages} {1009}
  (\bibinfo {year} {1998})},\ \Eprint {http://arxiv.org/abs/astro-ph/9805201}
  {arXiv:astro-ph/9805201 [astro-ph]} \BibitemShut {NoStop}%
\bibitem [{\citenamefont {Perlmutter}\ \emph {et~al.}(1999)\citenamefont
  {Perlmutter} \emph {et~al.}}]{Perlmutter:1998np}%
  \BibitemOpen
  \bibfield  {author} {\bibinfo {author} {\bibfnamefont {S.}~\bibnamefont
  {Perlmutter}} \emph {et~al.} (\bibinfo {collaboration} {Supernova Cosmology
  Project}),\ }\href {\doibase 10.1086/307221} {\bibfield  {journal} {\bibinfo
  {journal} {Astrophys.J.}\ }\textbf {\bibinfo {volume} {517}},\ \bibinfo
  {pages} {565} (\bibinfo {year} {1999})},\ \Eprint
  {http://arxiv.org/abs/astro-ph/9812133} {arXiv:astro-ph/9812133 [astro-ph]}
  \BibitemShut {NoStop}%
\bibitem [{\citenamefont {Clifton}\ \emph {et~al.}(2012)\citenamefont
  {Clifton}, \citenamefont {Ferreira}, \citenamefont {Padilla},\ and\
  \citenamefont {Skordis}}]{Clifton:2011jh}%
  \BibitemOpen
  \bibfield  {author} {\bibinfo {author} {\bibfnamefont {T.}~\bibnamefont
  {Clifton}}, \bibinfo {author} {\bibfnamefont {P.~G.}\ \bibnamefont
  {Ferreira}}, \bibinfo {author} {\bibfnamefont {A.}~\bibnamefont {Padilla}}, \
  and\ \bibinfo {author} {\bibfnamefont {C.}~\bibnamefont {Skordis}},\ }\href
  {\doibase 10.1016/j.physrep.2012.01.001} {\bibfield  {journal} {\bibinfo
  {journal} {Phys.Rept.}\ }\textbf {\bibinfo {volume} {513}},\ \bibinfo {pages}
  {1} (\bibinfo {year} {2012})},\ \Eprint {http://arxiv.org/abs/1106.2476}
  {arXiv:1106.2476 [astro-ph.CO]} \BibitemShut {NoStop}%
\bibitem [{\citenamefont {Joyce}\ \emph {et~al.}(2014)\citenamefont {Joyce},
  \citenamefont {Jain}, \citenamefont {Khoury},\ and\ \citenamefont
  {Trodden}}]{Joyce:2014kja}%
  \BibitemOpen
  \bibfield  {author} {\bibinfo {author} {\bibfnamefont {A.}~\bibnamefont
  {Joyce}}, \bibinfo {author} {\bibfnamefont {B.}~\bibnamefont {Jain}},
  \bibinfo {author} {\bibfnamefont {J.}~\bibnamefont {Khoury}}, \ and\ \bibinfo
  {author} {\bibfnamefont {M.}~\bibnamefont {Trodden}},\ }\href@noop {} {\
  (\bibinfo {year} {2014})},\ \Eprint {http://arxiv.org/abs/1407.0059}
  {arXiv:1407.0059 [astro-ph.CO]} \BibitemShut {NoStop}%
\bibitem [{\citenamefont {Khoury}\ and\ \citenamefont
  {Weltman}(2004{\natexlab{a}})}]{Khoury:2003aq}%
  \BibitemOpen
  \bibfield  {author} {\bibinfo {author} {\bibfnamefont {J.}~\bibnamefont
  {Khoury}}\ and\ \bibinfo {author} {\bibfnamefont {A.}~\bibnamefont
  {Weltman}},\ }\href {\doibase 10.1103/PhysRevLett.93.171104} {\bibfield
  {journal} {\bibinfo  {journal} {Phys.Rev.Lett.}\ }\textbf {\bibinfo {volume}
  {93}},\ \bibinfo {pages} {171104} (\bibinfo {year} {2004}{\natexlab{a}})},\
  \Eprint {http://arxiv.org/abs/astro-ph/0309300} {arXiv:astro-ph/0309300
  [astro-ph]} \BibitemShut {NoStop}%
\bibitem [{\citenamefont {Khoury}\ and\ \citenamefont
  {Weltman}(2004{\natexlab{b}})}]{Khoury:2003rn}%
  \BibitemOpen
  \bibfield  {author} {\bibinfo {author} {\bibfnamefont {J.}~\bibnamefont
  {Khoury}}\ and\ \bibinfo {author} {\bibfnamefont {A.}~\bibnamefont
  {Weltman}},\ }\href {\doibase 10.1103/PhysRevD.69.044026} {\bibfield
  {journal} {\bibinfo  {journal} {Phys. Rev.}\ }\textbf {\bibinfo {volume}
  {D69}},\ \bibinfo {pages} {044026} (\bibinfo {year} {2004}{\natexlab{b}})},\
  \Eprint {http://arxiv.org/abs/astro-ph/0309411} {arXiv:astro-ph/0309411}
  \BibitemShut {NoStop}%
\bibitem [{\citenamefont {Hinterbichler}\ \emph {et~al.}(2011)\citenamefont
  {Hinterbichler}, \citenamefont {Khoury}, \citenamefont {Levy},\ and\
  \citenamefont {Matas}}]{Hinterbichler:2011ca}%
  \BibitemOpen
  \bibfield  {author} {\bibinfo {author} {\bibfnamefont {K.}~\bibnamefont
  {Hinterbichler}}, \bibinfo {author} {\bibfnamefont {J.}~\bibnamefont
  {Khoury}}, \bibinfo {author} {\bibfnamefont {A.}~\bibnamefont {Levy}}, \ and\
  \bibinfo {author} {\bibfnamefont {A.}~\bibnamefont {Matas}},\ }\href@noop {}
  {\  (\bibinfo {year} {2011})},\ \Eprint {http://arxiv.org/abs/1107.2112}
  {arXiv:1107.2112 [astro-ph.CO]} \BibitemShut {NoStop}%
\bibitem [{\citenamefont {Brax}\ \emph {et~al.}(2010)\citenamefont {Brax},
  \citenamefont {van~de Bruck}, \citenamefont {Davis},\ and\ \citenamefont
  {Shaw}}]{Brax:2010gi}%
  \BibitemOpen
  \bibfield  {author} {\bibinfo {author} {\bibfnamefont {P.}~\bibnamefont
  {Brax}}, \bibinfo {author} {\bibfnamefont {C.}~\bibnamefont {van~de Bruck}},
  \bibinfo {author} {\bibfnamefont {A.-C.}\ \bibnamefont {Davis}}, \ and\
  \bibinfo {author} {\bibfnamefont {D.}~\bibnamefont {Shaw}},\ }\href {\doibase
  10.1103/PhysRevD.82.063519} {\bibfield  {journal} {\bibinfo  {journal} {Phys.
  Rev.}\ }\textbf {\bibinfo {volume} {D82}},\ \bibinfo {pages} {063519}
  (\bibinfo {year} {2010})},\ \Eprint {http://arxiv.org/abs/1005.3735}
  {arXiv:1005.3735 [astro-ph.CO]} \BibitemShut {NoStop}%
\bibitem [{\citenamefont {Vainshtein}(1972)}]{Vainshtein:1972sx}%
  \BibitemOpen
  \bibfield  {author} {\bibinfo {author} {\bibfnamefont {A.}~\bibnamefont
  {Vainshtein}},\ }\href {\doibase 10.1016/0370-2693(72)90147-5} {\bibfield
  {journal} {\bibinfo  {journal} {Phys.Lett.}\ }\textbf {\bibinfo {volume}
  {B39}},\ \bibinfo {pages} {393} (\bibinfo {year} {1972})}\BibitemShut
  {NoStop}%
\bibitem [{Note1()}]{Note1}%
  \BibitemOpen
  \bibinfo {note} {Note that some theories such as massive gravity \cite
  {deRham:2010kj} utilise these mechanisms, although the coupling to matter is
  only seen explicitly in the decoupling limit.}\BibitemShut {Stop}%
\bibitem [{\citenamefont {{Bekenstein}}(1992)}]{1992mgm..conf..905B}%
  \BibitemOpen
  \bibfield  {author} {\bibinfo {author} {\bibfnamefont {J.~D.}\ \bibnamefont
  {{Bekenstein}}},\ }in\ \href@noop {} {\emph {\bibinfo {booktitle} {Marcel
  Grossmann Meeting on General Relativity}}},\ \bibinfo {editor} {edited by\
  \bibinfo {editor} {\bibfnamefont {F.}~\bibnamefont {{Sat{\= o}}}}\ and\
  \bibinfo {editor} {\bibfnamefont {T.}~\bibnamefont {{Nakamura}}}}\ (\bibinfo
  {year} {1992})\ p.\ \bibinfo {pages} {905}\BibitemShut {NoStop}%
\bibitem [{\citenamefont {Bekenstein}(1993)}]{Bekenstein:1992pj}%
  \BibitemOpen
  \bibfield  {author} {\bibinfo {author} {\bibfnamefont {J.~D.}\ \bibnamefont
  {Bekenstein}},\ }\href {\doibase 10.1103/PhysRevD.48.3641} {\bibfield
  {journal} {\bibinfo  {journal} {Phys.Rev.}\ }\textbf {\bibinfo {volume}
  {D48}},\ \bibinfo {pages} {3641} (\bibinfo {year} {1993})},\ \Eprint
  {http://arxiv.org/abs/gr-qc/9211017} {arXiv:gr-qc/9211017 [gr-qc]}
  \BibitemShut {NoStop}%
\bibitem [{\citenamefont {Koivisto}(2008)}]{Koivisto:2008ak}%
  \BibitemOpen
  \bibfield  {author} {\bibinfo {author} {\bibfnamefont {T.~S.}\ \bibnamefont
  {Koivisto}},\ }\href@noop {} {\  (\bibinfo {year} {2008})},\ \Eprint
  {http://arxiv.org/abs/0811.1957} {arXiv:0811.1957 [astro-ph]} \BibitemShut
  {NoStop}%
\bibitem [{\citenamefont {Zumalacarregui}\ \emph {et~al.}(2010)\citenamefont
  {Zumalacarregui}, \citenamefont {Koivisto}, \citenamefont {Mota},\ and\
  \citenamefont {Ruiz-Lapuente}}]{Zumalacarregui:2010wj}%
  \BibitemOpen
  \bibfield  {author} {\bibinfo {author} {\bibfnamefont {M.}~\bibnamefont
  {Zumalacarregui}}, \bibinfo {author} {\bibfnamefont {T.}~\bibnamefont
  {Koivisto}}, \bibinfo {author} {\bibfnamefont {D.}~\bibnamefont {Mota}}, \
  and\ \bibinfo {author} {\bibfnamefont {P.}~\bibnamefont {Ruiz-Lapuente}},\
  }\href {\doibase 10.1088/1475-7516/2010/05/038} {\bibfield  {journal}
  {\bibinfo  {journal} {JCAP}\ }\textbf {\bibinfo {volume} {1005}},\ \bibinfo
  {pages} {038} (\bibinfo {year} {2010})},\ \Eprint
  {http://arxiv.org/abs/1004.2684} {arXiv:1004.2684 [astro-ph.CO]} \BibitemShut
  {NoStop}%
\bibitem [{\citenamefont {Noller}(2012)}]{Noller:2012sv}%
  \BibitemOpen
  \bibfield  {author} {\bibinfo {author} {\bibfnamefont {J.}~\bibnamefont
  {Noller}},\ }\href {\doibase 10.1088/1475-7516/2012/07/013} {\bibfield
  {journal} {\bibinfo  {journal} {JCAP}\ }\textbf {\bibinfo {volume} {1207}},\
  \bibinfo {pages} {013} (\bibinfo {year} {2012})},\ \Eprint
  {http://arxiv.org/abs/1203.6639} {arXiv:1203.6639 [gr-qc]} \BibitemShut
  {NoStop}%
\bibitem [{\citenamefont {Zumalacarregui}\ \emph {et~al.}(2013)\citenamefont
  {Zumalacarregui}, \citenamefont {Koivisto},\ and\ \citenamefont
  {Mota}}]{Zumalacarregui:2012us}%
  \BibitemOpen
  \bibfield  {author} {\bibinfo {author} {\bibfnamefont {M.}~\bibnamefont
  {Zumalacarregui}}, \bibinfo {author} {\bibfnamefont {T.~S.}\ \bibnamefont
  {Koivisto}}, \ and\ \bibinfo {author} {\bibfnamefont {D.~F.}\ \bibnamefont
  {Mota}},\ }\href {\doibase 10.1103/PhysRevD.87.083010} {\bibfield  {journal}
  {\bibinfo  {journal} {Phys.Rev.}\ }\textbf {\bibinfo {volume} {D87}},\
  \bibinfo {pages} {083010} (\bibinfo {year} {2013})},\ \Eprint
  {http://arxiv.org/abs/1210.8016} {arXiv:1210.8016 [astro-ph.CO]} \BibitemShut
  {NoStop}%
\bibitem [{\citenamefont {Zumalacárregui}\ and\ \citenamefont
  {García-Bellido}(2014)}]{Zumalacarregui:2013pma}%
  \BibitemOpen
  \bibfield  {author} {\bibinfo {author} {\bibfnamefont {M.}~\bibnamefont
  {Zumalacárregui}}\ and\ \bibinfo {author} {\bibfnamefont {J.}~\bibnamefont
  {García-Bellido}},\ }\href {\doibase 10.1103/PhysRevD.89.064046} {\bibfield
  {journal} {\bibinfo  {journal} {Phys.Rev.}\ }\textbf {\bibinfo {volume}
  {D89}},\ \bibinfo {pages} {064046} (\bibinfo {year} {2014})},\ \Eprint
  {http://arxiv.org/abs/1308.4685} {arXiv:1308.4685 [gr-qc]} \BibitemShut
  {NoStop}%
\bibitem [{\citenamefont {Koivisto}\ \emph {et~al.}(2013)\citenamefont
  {Koivisto}, \citenamefont {Wills},\ and\ \citenamefont
  {Zavala}}]{Koivisto:2013fta}%
  \BibitemOpen
  \bibfield  {author} {\bibinfo {author} {\bibfnamefont {T.}~\bibnamefont
  {Koivisto}}, \bibinfo {author} {\bibfnamefont {D.}~\bibnamefont {Wills}}, \
  and\ \bibinfo {author} {\bibfnamefont {I.}~\bibnamefont {Zavala}},\
  }\href@noop {} {\  (\bibinfo {year} {2013})},\ \Eprint
  {http://arxiv.org/abs/1312.2597} {arXiv:1312.2597 [hep-th]} \BibitemShut
  {NoStop}%
\bibitem [{\citenamefont {Koivisto}\ and\ \citenamefont
  {Urban}(2014)}]{Koivisto:2014gia}%
  \BibitemOpen
  \bibfield  {author} {\bibinfo {author} {\bibfnamefont {T.~S.}\ \bibnamefont
  {Koivisto}}\ and\ \bibinfo {author} {\bibfnamefont {F.~R.}\ \bibnamefont
  {Urban}},\ }\href@noop {} {\  (\bibinfo {year} {2014})},\ \Eprint
  {http://arxiv.org/abs/1407.3445} {arXiv:1407.3445 [astro-ph.CO]} \BibitemShut
  {NoStop}%
\bibitem [{\citenamefont {Sakstein}(2014)}]{Sakstein:2014isa}%
  \BibitemOpen
  \bibfield  {author} {\bibinfo {author} {\bibfnamefont {J.}~\bibnamefont
  {Sakstein}},\ }\href@noop {} {\  (\bibinfo {year} {2014})},\ \Eprint
  {http://arxiv.org/abs/1409.1734} {arXiv:1409.1734 [astro-ph.CO]} \BibitemShut
  {NoStop}%
\bibitem [{\citenamefont {Kaloper}(2004)}]{Kaloper:2003yf}%
  \BibitemOpen
  \bibfield  {author} {\bibinfo {author} {\bibfnamefont {N.}~\bibnamefont
  {Kaloper}},\ }\href {\doibase 10.1016/j.physletb.2004.01.005} {\bibfield
  {journal} {\bibinfo  {journal} {Phys.Lett.}\ }\textbf {\bibinfo {volume}
  {B583}},\ \bibinfo {pages} {1} (\bibinfo {year} {2004})},\ \Eprint
  {http://arxiv.org/abs/hep-ph/0312002} {arXiv:hep-ph/0312002 [hep-ph]}
  \BibitemShut {NoStop}%
\bibitem [{\citenamefont {Davis}\ \emph {et~al.}(2012)\citenamefont {Davis},
  \citenamefont {Lim}, \citenamefont {Sakstein},\ and\ \citenamefont
  {Shaw}}]{Davis:2011qf}%
  \BibitemOpen
  \bibfield  {author} {\bibinfo {author} {\bibfnamefont {A.-C.}\ \bibnamefont
  {Davis}}, \bibinfo {author} {\bibfnamefont {E.~A.}\ \bibnamefont {Lim}},
  \bibinfo {author} {\bibfnamefont {J.}~\bibnamefont {Sakstein}}, \ and\
  \bibinfo {author} {\bibfnamefont {D.}~\bibnamefont {Shaw}},\ }\href {\doibase
  10.1103/PhysRevD.85.123006} {\bibfield  {journal} {\bibinfo  {journal}
  {Phys.Rev.}\ }\textbf {\bibinfo {volume} {D85}},\ \bibinfo {pages} {123006}
  (\bibinfo {year} {2012})},\ \Eprint {http://arxiv.org/abs/1102.5278}
  {arXiv:1102.5278 [astro-ph.CO]} \BibitemShut {NoStop}%
\bibitem [{\citenamefont {Jain}\ \emph {et~al.}(2013)\citenamefont {Jain},
  \citenamefont {Vikram},\ and\ \citenamefont {Sakstein}}]{Jain:2012tn}%
  \BibitemOpen
  \bibfield  {author} {\bibinfo {author} {\bibfnamefont {B.}~\bibnamefont
  {Jain}}, \bibinfo {author} {\bibfnamefont {V.}~\bibnamefont {Vikram}}, \ and\
  \bibinfo {author} {\bibfnamefont {J.}~\bibnamefont {Sakstein}},\ }\href
  {\doibase 10.1088/0004-637X/779/1/39} {\bibfield  {journal} {\bibinfo
  {journal} {Astrophys.J.}\ }\textbf {\bibinfo {volume} {779}},\ \bibinfo
  {pages} {39} (\bibinfo {year} {2013})},\ \Eprint
  {http://arxiv.org/abs/1204.6044} {arXiv:1204.6044 [astro-ph.CO]} \BibitemShut
  {NoStop}%
\bibitem [{\citenamefont {Brax}\ \emph
  {et~al.}(2013{\natexlab{a}})\citenamefont {Brax}, \citenamefont {Davis},\
  and\ \citenamefont {Sakstein}}]{Brax:2013uh}%
  \BibitemOpen
  \bibfield  {author} {\bibinfo {author} {\bibfnamefont {P.}~\bibnamefont
  {Brax}}, \bibinfo {author} {\bibfnamefont {A.-C.}\ \bibnamefont {Davis}}, \
  and\ \bibinfo {author} {\bibfnamefont {J.}~\bibnamefont {Sakstein}},\
  }\href@noop {} {\  (\bibinfo {year} {2013}{\natexlab{a}})},\ \Eprint
  {http://arxiv.org/abs/1301.5587} {arXiv:1301.5587 [gr-qc]} \BibitemShut
  {NoStop}%
\bibitem [{\citenamefont {Sakstein}(2013)}]{Sakstein:2013pda}%
  \BibitemOpen
  \bibfield  {author} {\bibinfo {author} {\bibfnamefont {J.}~\bibnamefont
  {Sakstein}},\ }\href {\doibase 10.1103/PhysRevD.88.124013} {\bibfield
  {journal} {\bibinfo  {journal} {Phys.Rev.}\ }\textbf {\bibinfo {volume}
  {D88}},\ \bibinfo {pages} {124013} (\bibinfo {year} {2013})},\ \Eprint
  {http://arxiv.org/abs/1309.0495} {arXiv:1309.0495 [astro-ph.CO]} \BibitemShut
  {NoStop}%
\bibitem [{\citenamefont {Vikram}\ \emph {et~al.}(2014)\citenamefont {Vikram},
  \citenamefont {Sakstein}, \citenamefont {Davis},\ and\ \citenamefont
  {Neil}}]{Vikram:2014uza}%
  \BibitemOpen
  \bibfield  {author} {\bibinfo {author} {\bibfnamefont {V.}~\bibnamefont
  {Vikram}}, \bibinfo {author} {\bibfnamefont {J.}~\bibnamefont {Sakstein}},
  \bibinfo {author} {\bibfnamefont {C.}~\bibnamefont {Davis}}, \ and\ \bibinfo
  {author} {\bibfnamefont {A.}~\bibnamefont {Neil}},\ }\href@noop {} {\
  (\bibinfo {year} {2014})},\ \Eprint {http://arxiv.org/abs/1407.6044}
  {arXiv:1407.6044 [astro-ph.CO]} \BibitemShut {NoStop}%
\bibitem [{\citenamefont {Sakstein}\ \emph {et~al.}(2014)\citenamefont
  {Sakstein}, \citenamefont {Jain},\ and\ \citenamefont
  {Vikram}}]{Sakstein:2014nfa}%
  \BibitemOpen
  \bibfield  {author} {\bibinfo {author} {\bibfnamefont {J.}~\bibnamefont
  {Sakstein}}, \bibinfo {author} {\bibfnamefont {B.}~\bibnamefont {Jain}}, \
  and\ \bibinfo {author} {\bibfnamefont {V.}~\bibnamefont {Vikram}},\ }\href
  {\doibase 10.1142/S0218271814420024} {\  (\bibinfo {year} {2014}),\
  10.1142/S0218271814420024},\ \Eprint {http://arxiv.org/abs/1409.3708}
  {arXiv:1409.3708 [astro-ph.CO]} \BibitemShut {NoStop}%
\bibitem [{\citenamefont {Freire}\ \emph {et~al.}(2012)\citenamefont {Freire},
  \citenamefont {Wex}, \citenamefont {Esposito-Farese}, \citenamefont
  {Verbiest}, \citenamefont {Bailes} \emph {et~al.}}]{Freire:2012mg}%
  \BibitemOpen
  \bibfield  {author} {\bibinfo {author} {\bibfnamefont {P.~C.}\ \bibnamefont
  {Freire}}, \bibinfo {author} {\bibfnamefont {N.}~\bibnamefont {Wex}},
  \bibinfo {author} {\bibfnamefont {G.}~\bibnamefont {Esposito-Farese}},
  \bibinfo {author} {\bibfnamefont {J.~P.}\ \bibnamefont {Verbiest}}, \bibinfo
  {author} {\bibfnamefont {M.}~\bibnamefont {Bailes}},  \emph {et~al.},\ }\href
  {\doibase 10.1111/j.1365-2966.2012.21253.x} {\bibfield  {journal} {\bibinfo
  {journal} {Mon.Not.Roy.Astron.Soc.}\ }\textbf {\bibinfo {volume} {423}},\
  \bibinfo {pages} {3328} (\bibinfo {year} {2012})},\ \Eprint
  {http://arxiv.org/abs/1205.1450} {arXiv:1205.1450 [astro-ph.GA]} \BibitemShut
  {NoStop}%
\bibitem [{\citenamefont {Khoury}(2014)}]{Khoury:2014tka}%
  \BibitemOpen
  \bibfield  {author} {\bibinfo {author} {\bibfnamefont {J.}~\bibnamefont
  {Khoury}},\ }\href@noop {} {\  (\bibinfo {year} {2014})},\ \Eprint
  {http://arxiv.org/abs/1409.0012} {arXiv:1409.0012 [hep-th]} \BibitemShut
  {NoStop}%
\bibitem [{Note2()}]{Note2}%
  \BibitemOpen
  \bibinfo {note} {One could screen this using the Damour-Polyakov effect \cite
  {Damour:1994zq}, although this has not yet been investigated.}\BibitemShut
  {Stop}%
\bibitem [{\citenamefont {Copeland}\ \emph {et~al.}(1998)\citenamefont
  {Copeland}, \citenamefont {Liddle},\ and\ \citenamefont
  {Wands}}]{Copeland:1997et}%
  \BibitemOpen
  \bibfield  {author} {\bibinfo {author} {\bibfnamefont {E.~J.}\ \bibnamefont
  {Copeland}}, \bibinfo {author} {\bibfnamefont {A.~R.}\ \bibnamefont
  {Liddle}}, \ and\ \bibinfo {author} {\bibfnamefont {D.}~\bibnamefont
  {Wands}},\ }\href {\doibase 10.1103/PhysRevD.57.4686} {\bibfield  {journal}
  {\bibinfo  {journal} {Phys.Rev.}\ }\textbf {\bibinfo {volume} {D57}},\
  \bibinfo {pages} {4686} (\bibinfo {year} {1998})},\ \Eprint
  {http://arxiv.org/abs/gr-qc/9711068} {arXiv:gr-qc/9711068 [gr-qc]}
  \BibitemShut {NoStop}%
\bibitem [{\citenamefont {Holden}\ and\ \citenamefont
  {Wands}(2000)}]{Holden:1999hm}%
  \BibitemOpen
  \bibfield  {author} {\bibinfo {author} {\bibfnamefont {D.~J.}\ \bibnamefont
  {Holden}}\ and\ \bibinfo {author} {\bibfnamefont {D.}~\bibnamefont {Wands}},\
  }\href {\doibase 10.1103/PhysRevD.61.043506} {\bibfield  {journal} {\bibinfo
  {journal} {Phys.Rev.}\ }\textbf {\bibinfo {volume} {D61}},\ \bibinfo {pages}
  {043506} (\bibinfo {year} {2000})},\ \Eprint
  {http://arxiv.org/abs/gr-qc/9908026} {arXiv:gr-qc/9908026 [gr-qc]}
  \BibitemShut {NoStop}%
\bibitem [{\citenamefont {Gumjudpai}\ \emph {et~al.}(2005)\citenamefont
  {Gumjudpai}, \citenamefont {Naskar}, \citenamefont {Sami},\ and\
  \citenamefont {Tsujikawa}}]{Gumjudpai:2005ry}%
  \BibitemOpen
  \bibfield  {author} {\bibinfo {author} {\bibfnamefont {B.}~\bibnamefont
  {Gumjudpai}}, \bibinfo {author} {\bibfnamefont {T.}~\bibnamefont {Naskar}},
  \bibinfo {author} {\bibfnamefont {M.}~\bibnamefont {Sami}}, \ and\ \bibinfo
  {author} {\bibfnamefont {S.}~\bibnamefont {Tsujikawa}},\ }\href {\doibase
  10.1088/1475-7516/2005/06/007} {\bibfield  {journal} {\bibinfo  {journal}
  {JCAP}\ }\textbf {\bibinfo {volume} {0506}},\ \bibinfo {pages} {007}
  (\bibinfo {year} {2005})},\ \Eprint {http://arxiv.org/abs/hep-th/0502191}
  {arXiv:hep-th/0502191 [hep-th]} \BibitemShut {NoStop}%
\bibitem [{\citenamefont {Copeland}\ \emph {et~al.}(2006)\citenamefont
  {Copeland}, \citenamefont {Sami},\ and\ \citenamefont
  {Tsujikawa}}]{Copeland:2006wr}%
  \BibitemOpen
  \bibfield  {author} {\bibinfo {author} {\bibfnamefont {E.~J.}\ \bibnamefont
  {Copeland}}, \bibinfo {author} {\bibfnamefont {M.}~\bibnamefont {Sami}}, \
  and\ \bibinfo {author} {\bibfnamefont {S.}~\bibnamefont {Tsujikawa}},\ }\href
  {\doibase 10.1142/S021827180600942X} {\bibfield  {journal} {\bibinfo
  {journal} {Int.J.Mod.Phys.}\ }\textbf {\bibinfo {volume} {D15}},\ \bibinfo
  {pages} {1753} (\bibinfo {year} {2006})},\ \Eprint
  {http://arxiv.org/abs/hep-th/0603057} {arXiv:hep-th/0603057 [hep-th]}
  \BibitemShut {NoStop}%
\bibitem [{\citenamefont {Hinshaw}\ \emph {et~al.}(2013)\citenamefont {Hinshaw}
  \emph {et~al.}}]{Hinshaw:2012aka}%
  \BibitemOpen
  \bibfield  {author} {\bibinfo {author} {\bibfnamefont {G.}~\bibnamefont
  {Hinshaw}} \emph {et~al.} (\bibinfo {collaboration} {WMAP}),\ }\href
  {\doibase 10.1088/0067-0049/208/2/19} {\bibfield  {journal} {\bibinfo
  {journal} {Astrophys.J.Suppl.}\ }\textbf {\bibinfo {volume} {208}},\ \bibinfo
  {pages} {19} (\bibinfo {year} {2013})},\ \Eprint
  {http://arxiv.org/abs/1212.5226} {arXiv:1212.5226 [astro-ph.CO]} \BibitemShut
  {NoStop}%
\bibitem [{Note3()}]{Note3}%
  \BibitemOpen
  \bibinfo {note} {Note that we are using the conventions of \cite
  {Sakstein:2014isa}. A dictionary to convert these conventions to others in
  the literature such as \cite {Zumalacarregui:2012us} can be found
  there.}\BibitemShut {Stop}%
\bibitem [{Note4()}]{Note4}%
  \BibitemOpen
  \bibinfo {note} {It is the Jordan frame energy-momentum tensor that is
  conserved since matter is minimally coupled in this frame.}\BibitemShut
  {Stop}%
\bibitem [{Note5()}]{Note5}%
  \BibitemOpen
  \bibinfo {note} {It is assumed that $\Phi \sim \Psi \sim \varphi \sim \Phi
  _{\protect \rm N}$, where $\Phi _{\protect \rm N}$ is the Newtonian
  potential, so that terms such as $\protect \mathaccentV {dot}05F{\varphi
  }^2\sim \Phi _{\protect \rm N}\nabla ^2\Phi _{\protect \rm N}$ can be
  neglected. This ansatz can be checked self-consistently (see the discussion
  in \cite {Sakstein:2014isa}) and allows one to systematically construct the
  non-relativistic limit.}\BibitemShut {Stop}%
\bibitem [{Note6()}]{Note6}%
  \BibitemOpen
  \bibinfo {note} {Note that this assumes $\protect \mathaccentV {dot}05F{\phi
  }_\infty \ll \Lambda $. If this is not the case then there are time-dependent
  corrections to Newton's constant (see the discussion below).}\BibitemShut
  {Stop}%
\bibitem [{Note7()}]{Note7}%
  \BibitemOpen
  \bibinfo {note} {One could perform the same analysis for the general model
  but this is cumbersome and does not add any more insight. The pathologies
  discussed here arise due to the disformal coupling and so it suffices to
  consider the simplest case for the purposes of elucidation.}\BibitemShut
  {Stop}%
\bibitem [{\citenamefont {Bellini}\ and\ \citenamefont
  {Sawicki}(2014)}]{Bellini:2014fua}%
  \BibitemOpen
  \bibfield  {author} {\bibinfo {author} {\bibfnamefont {E.}~\bibnamefont
  {Bellini}}\ and\ \bibinfo {author} {\bibfnamefont {I.}~\bibnamefont
  {Sawicki}},\ }\href {\doibase 10.1088/1475-7516/2014/07/050} {\bibfield
  {journal} {\bibinfo  {journal} {JCAP}\ }\textbf {\bibinfo {volume} {1407}},\
  \bibinfo {pages} {050} (\bibinfo {year} {2014})},\ \Eprint
  {http://arxiv.org/abs/1404.3713} {arXiv:1404.3713 [astro-ph.CO]} \BibitemShut
  {NoStop}%
\bibitem [{\citenamefont {van~de Bruck}\ and\ \citenamefont
  {Sculthorpe}(2013)}]{vandeBruck:2012vq}%
  \BibitemOpen
  \bibfield  {author} {\bibinfo {author} {\bibfnamefont {C.}~\bibnamefont
  {van~de Bruck}}\ and\ \bibinfo {author} {\bibfnamefont {G.}~\bibnamefont
  {Sculthorpe}},\ }\href {\doibase 10.1103/PhysRevD.87.044004} {\bibfield
  {journal} {\bibinfo  {journal} {Phys.Rev.}\ }\textbf {\bibinfo {volume}
  {D87}},\ \bibinfo {pages} {044004} (\bibinfo {year} {2013})},\ \Eprint
  {http://arxiv.org/abs/1210.2168} {arXiv:1210.2168 [astro-ph.CO]} \BibitemShut
  {NoStop}%
\bibitem [{\citenamefont {Brax}\ \emph
  {et~al.}(2013{\natexlab{b}})\citenamefont {Brax}, \citenamefont {Burrage},
  \citenamefont {Davis},\ and\ \citenamefont {Gubitosi}}]{Brax:2013nsa}%
  \BibitemOpen
  \bibfield  {author} {\bibinfo {author} {\bibfnamefont {P.}~\bibnamefont
  {Brax}}, \bibinfo {author} {\bibfnamefont {C.}~\bibnamefont {Burrage}},
  \bibinfo {author} {\bibfnamefont {A.-C.}\ \bibnamefont {Davis}}, \ and\
  \bibinfo {author} {\bibfnamefont {G.}~\bibnamefont {Gubitosi}},\ }\href
  {\doibase 10.1088/1475-7516/2013/11/001} {\bibfield  {journal} {\bibinfo
  {journal} {JCAP}\ }\textbf {\bibinfo {volume} {1311}},\ \bibinfo {pages}
  {001} (\bibinfo {year} {2013}{\natexlab{b}})},\ \Eprint
  {http://arxiv.org/abs/1306.4168} {arXiv:1306.4168 [astro-ph.CO]} \BibitemShut
  {NoStop}%
\bibitem [{\citenamefont {van~de Bruck}\ \emph {et~al.}(2013)\citenamefont
  {van~de Bruck}, \citenamefont {Morrice},\ and\ \citenamefont
  {Vu}}]{vandeBruck:2013yxa}%
  \BibitemOpen
  \bibfield  {author} {\bibinfo {author} {\bibfnamefont {C.}~\bibnamefont
  {van~de Bruck}}, \bibinfo {author} {\bibfnamefont {J.}~\bibnamefont
  {Morrice}}, \ and\ \bibinfo {author} {\bibfnamefont {S.}~\bibnamefont {Vu}},\
  }\href {\doibase 10.1103/PhysRevLett.111.161302} {\bibfield  {journal}
  {\bibinfo  {journal} {Phys.Rev.Lett.}\ }\textbf {\bibinfo {volume} {111}},\
  \bibinfo {pages} {161302} (\bibinfo {year} {2013})},\ \Eprint
  {http://arxiv.org/abs/1303.1773} {arXiv:1303.1773 [astro-ph.CO]} \BibitemShut
  {NoStop}%
\bibitem [{\citenamefont {Bettoni}\ and\ \citenamefont
  {Liberati}(2013)}]{Bettoni:2013diz}%
  \BibitemOpen
  \bibfield  {author} {\bibinfo {author} {\bibfnamefont {D.}~\bibnamefont
  {Bettoni}}\ and\ \bibinfo {author} {\bibfnamefont {S.}~\bibnamefont
  {Liberati}},\ }\href {\doibase 10.1103/PhysRevD.88.084020} {\bibfield
  {journal} {\bibinfo  {journal} {Phys.Rev.}\ }\textbf {\bibinfo {volume}
  {D88}},\ \bibinfo {pages} {084020} (\bibinfo {year} {2013})},\ \Eprint
  {http://arxiv.org/abs/1306.6724} {arXiv:1306.6724 [gr-qc]} \BibitemShut
  {NoStop}%
\bibitem [{\citenamefont {Brax}(2012)}]{Brax:2012yi}%
  \BibitemOpen
  \bibfield  {author} {\bibinfo {author} {\bibfnamefont {P.}~\bibnamefont
  {Brax}},\ }\href@noop {} {\  (\bibinfo {year} {2012})},\ \Eprint
  {http://arxiv.org/abs/1211.5237} {arXiv:1211.5237 [hep-th]} \BibitemShut
  {NoStop}%
\bibitem [{\citenamefont {Barreira}\ \emph {et~al.}(2013)\citenamefont
  {Barreira}, \citenamefont {Li}, \citenamefont {Baugh},\ and\ \citenamefont
  {Pascoli}}]{Barreira:2013xea}%
  \BibitemOpen
  \bibfield  {author} {\bibinfo {author} {\bibfnamefont {A.}~\bibnamefont
  {Barreira}}, \bibinfo {author} {\bibfnamefont {B.}~\bibnamefont {Li}},
  \bibinfo {author} {\bibfnamefont {C.}~\bibnamefont {Baugh}}, \ and\ \bibinfo
  {author} {\bibfnamefont {S.}~\bibnamefont {Pascoli}},\ }\href@noop {} {\
  (\bibinfo {year} {2013})},\ \Eprint {http://arxiv.org/abs/1308.3699}
  {arXiv:1308.3699 [astro-ph.CO]} \BibitemShut {NoStop}%
\bibitem [{Note8()}]{Note8}%
  \BibitemOpen
  \bibinfo {note} {This is because the analysis employed below uses $\protect
  \qopname \relax o{ln}a(t)$ as a time coordinate and eliminates $H(a)$ through
  the Friedmann constraint. This choice implicitly assumes that $a(t)$ is
  monotonically increasing and can hence be used as a time coordinate. One
  could instead work with a larger phase space in order to treat $H$ as a
  dynamical variable and capture any deviations from monotonicity but it is
  simpler to work in the Jordan frame. This exercise is postponed for follow-up
  work.}\BibitemShut {Stop}%
\bibitem [{Note9()}]{Note9}%
  \BibitemOpen
  \bibinfo {note} {One may wonder whether re-introducing $A$ and $B$ can change
  this argument. One can always perform a field re-definition to remove the
  function multiplying the disformal factor and hence any effects of having
  $B\not =1$ can be equivalently thought of as arising due to non-canonical
  kinetic terms for the scalar. The conformal factor is less subtle but in this
  case any acceleration in the Jordan frame is due to conformal effects only
  and hence this scenario is not relevant for this work.}\BibitemShut {Stop}%
\bibitem [{Note10()}]{Note10}%
  \BibitemOpen
  \bibinfo {note} {Note that this was studied briefly in \cite
  {Zumalacarregui:2012us}, Appendix C.}\BibitemShut {Stop}%
\bibitem [{Note11()}]{Note11}%
  \BibitemOpen
  \bibinfo {note} {By this, we mean that at any fixed time the fixed points
  found here will be solutions of the equations. Whether or not they are
  reached however depends on the specific model and in general one expects new
  fixed points and that the properties found here may be destroyed. In
  practice, models where the dynamics are predictable---for example potentials
  with minima so that we expect $\lambda (\phi )\rightarrow 0$ at late
  times---tend to approach these fixed points but we stress that a more
  complicated analysis is required to draw any definite conclusions. This is
  especially true in cases where either the potential or the disformal factor
  can become zero.}\BibitemShut {Stop}%
\bibitem [{Note12()}]{Note12}%
  \BibitemOpen
  \bibinfo {note} {The parameter space is spanned by $\gamma $, $\alpha $,
  $\lambda $ and $\Lambda /m_0$}\BibitemShut {NoStop}%
\bibitem [{Note13()}]{Note13}%
  \BibitemOpen
  \bibinfo {note} {One may solve for $\phi $ from the definition of $y$ to find
  $B=(\protect \sqrt {3}Hy)^{2\gamma /\lambda }$. When $\gamma =\lambda /2$ one
  can scale $H$ out of the equations without the need to define a new quantity
  $z$ whereas when this is not the case a new variable is needed. Therefore,
  one may think of this tuning as an enhanced symmetry of the equations whereby
  one can scale $H$, $V$ and $\protect \mathaccentV {dot}05F{\phi }_\infty $ to
  obtain new solutions of the system without the need to scale
  $B$.}\BibitemShut {Stop}%
\bibitem [{\citenamefont {Amendola}(2000)}]{Amendola:1999er}%
  \BibitemOpen
  \bibfield  {author} {\bibinfo {author} {\bibfnamefont {L.}~\bibnamefont
  {Amendola}},\ }\href {\doibase 10.1103/PhysRevD.62.043511} {\bibfield
  {journal} {\bibinfo  {journal} {Phys.Rev.}\ }\textbf {\bibinfo {volume}
  {D62}},\ \bibinfo {pages} {043511} (\bibinfo {year} {2000})},\ \Eprint
  {http://arxiv.org/abs/astro-ph/9908023} {arXiv:astro-ph/9908023 [astro-ph]}
  \BibitemShut {NoStop}%
\bibitem [{Note14()}]{Note14}%
  \BibitemOpen
  \bibinfo {note} {Note that $z=0$ does not necessarily mean that $B=0$. Since
  $z\propto H$ these points may correspond to the infinite future where
  $H=0$.}\BibitemShut {Stop}%
\bibitem [{Note15()}]{Note15}%
  \BibitemOpen
  \bibinfo {note} {We have not included fixed points where $Z>1$ or $Z<0$,
  which lie outside the physical state space. Furthermore, one finds that there
  is an additional fixed point when $\gamma \rightarrow \infty $ where $Z=1$,
  which we also ignore. This may be relevant for models where $\gamma $ can
  reach infinity.}\BibitemShut {Stop}%
\bibitem [{\citenamefont {Alho}\ and\ \citenamefont
  {Uggla}(2014)}]{Alho:2014fha}%
  \BibitemOpen
  \bibfield  {author} {\bibinfo {author} {\bibfnamefont {A.}~\bibnamefont
  {Alho}}\ and\ \bibinfo {author} {\bibfnamefont {C.}~\bibnamefont {Uggla}},\
  }\href@noop {} {\  (\bibinfo {year} {2014})},\ \Eprint
  {http://arxiv.org/abs/1406.0438} {arXiv:1406.0438 [gr-qc]} \BibitemShut
  {NoStop}%
\bibitem [{Note16()}]{Note16}%
  \BibitemOpen
  \bibinfo {note} {One should note that we refer to this phenomena as \protect
  \textit {pathology resistance} in order to make contact with previous works.
  We consider models exhibiting metric singularities to be unviable for the
  reasons discussed in section \ref {sec:via}. One should note that it is
  possible to construct theories where the visible sector coupled to the
  Einstein frame metric but dark matter couples to the Jordan frame metric. In
  this case, it is less clear whether these solutions are indeed pathological
  since the non-relativistic limit is well-defined for visible matter and
  fifth-forces are not an issue.}\BibitemShut {Stop}%
\bibitem [{Note17()}]{Note17}%
  \BibitemOpen
  \bibinfo {note} {Note that this data set assumes that $\omega $ is fixed
  whereas it varies in our model and so a more realistic method would to use a
  varying $\omega $ fit such as the $\omega _0$--$\omega _a$ parametrisation.
  The analysis here is a proof of principle only and so we will not concern
  ourselves with a more realistic data analysis.}\BibitemShut {Stop}%
\bibitem [{\citenamefont {de~Rham}\ \emph {et~al.}(2011)\citenamefont
  {de~Rham}, \citenamefont {Gabadadze},\ and\ \citenamefont
  {Tolley}}]{deRham:2010kj}%
  \BibitemOpen
  \bibfield  {author} {\bibinfo {author} {\bibfnamefont {C.}~\bibnamefont
  {de~Rham}}, \bibinfo {author} {\bibfnamefont {G.}~\bibnamefont {Gabadadze}},
  \ and\ \bibinfo {author} {\bibfnamefont {A.~J.}\ \bibnamefont {Tolley}},\
  }\href {\doibase 10.1103/PhysRevLett.106.231101} {\bibfield  {journal}
  {\bibinfo  {journal} {Phys.Rev.Lett.}\ }\textbf {\bibinfo {volume} {106}},\
  \bibinfo {pages} {231101} (\bibinfo {year} {2011})},\ \Eprint
  {http://arxiv.org/abs/1011.1232} {arXiv:1011.1232 [hep-th]} \BibitemShut
  {NoStop}%
\bibitem [{\citenamefont {Damour}\ and\ \citenamefont
  {Polyakov}(1994)}]{Damour:1994zq}%
  \BibitemOpen
  \bibfield  {author} {\bibinfo {author} {\bibfnamefont {T.}~\bibnamefont
  {Damour}}\ and\ \bibinfo {author} {\bibfnamefont {A.~M.}\ \bibnamefont
  {Polyakov}},\ }\href {\doibase 10.1016/0550-3213(94)90143-0} {\bibfield
  {journal} {\bibinfo  {journal} {Nucl. Phys.}\ }\textbf {\bibinfo {volume}
  {B423}},\ \bibinfo {pages} {532} (\bibinfo {year} {1994})},\ \Eprint
  {http://arxiv.org/abs/hep-th/9401069} {arXiv:hep-th/9401069} \BibitemShut
  {NoStop}%
\bibitem [{Note18()}]{Note18}%
  \BibitemOpen
  \bibinfo {note} {We are interested in stable fixed points since these
  describe the late-time behaviour of the system and so we will not consider
  positive eigenvalues here.}\BibitemShut {Stop}%
\end{thebibliography}%

\appendix

\section{Dynamical Systems}\label{sec:dyna}

In this section we briefly review the aspects of dynamical systems theory required to study the disformal system. 

\subsection{Fixed Points and Stability}

Consider a system described by $n$ first-order ordinary differential equations for $n$ variables $X_i$ as a function of some ``time'' coordinate $t$ 
and let a dot denote derivatives with respect to this. If the system can be written in the form
\begin{equation}
 \frac{\dd X_i}{\dd t}=f_i(\{X_j\}),
\end{equation}
then it is known as an \textit{autonomous} system and one can use dynamical systems techniques to classify the solutions. The phase space of 
the system is the $n$-dimensional space spanned by $\{X_j\}$ and solutions of the system correspond to trajectories in this space. A fixed point of 
the system is one where $f_i=0\,\forall i$. This is a set of $n$ algebraic equations that can be solved for the values of the fixed points 
$\{X_j^{\rm c}\}$. If one tunes the variables to $\{X_j^{\rm c}\}$ then the system will not evolve but one is generally interested in the behaviour of 
arbitrary trajectories. In particular, if the fixed points are such that trajectories flow towards them at late times then the final state of the 
system is known independently of the initial conditions and so one can make important inferences about the late-time behaviour of all solutions. In 
this case, the point is known as an \textit{attractor}. If the trajectories flow away from the fixed point it is known as a \textit{repellor}. 
Finding and classifying all of the fixed points of a system is tantamount to understanding the behaviour of all possible solutions, which is 
especially important if one lacks analytic solutions. Cosmologically, one can calculate several important quantities such as the density parameter 
and the equation of state and so dynamical systems techniques have become an important tool to assess the viability of dark energy models.

One may determine the stability of the fixed points as follows. Consider linearising the system about a fixed point such that $X_i=X_i^{\rm c}+\delta 
X_i$. $\delta X_i$ satisfies the equation 
\begin{equation}
\dot{ \delta X}_i=M_{ij}\delta X_i
\end{equation}
where
\begin{equation}\label{eq:matdef}
 M_{ij}=\frac{\partial f_i}{\partial X_j},
\end{equation}
and we have ignored second order contributions since $\delta X_i$ is a small perturbation. This means that the stability analysis holds for regions 
of phase space close enough to the fixed points such that this linearisation is a good approximation. Changing to the 
eigenbasis $\vec{e}_i$ of $M$ one has ($A_j$ are new variables parametrising the system in the eigenbasis)
\begin{equation}
 X_i=\sum_j A_j(\vec{e}_j)_i
\end{equation}
so that the equation for $\delta A_j$ is
\begin{equation}
 \dot{\delta A}_j=e_j\delta A_j,
\end{equation}
where $e_j$ is the eigenvalue associated with $\vec{e}_j$. One then has $\delta A_j\sim e^{e_j t}$ and so the stability of the fixed point in the 
direction $\vec{e}_j$ is determined by $e_j$. There are several possibilities:
\begin{itemize}
 \item $e_j$ is real and $e_j>0\,\forall j$: Trajectories in the direction flow away from the fixed point and it is known as an \textit{unstable 
node}.
 \item $e_j$ is real and $e_j<0\,\forall j$: Trajectories in the direction flow towards the fixed point and it is known as an \textit{stable 
node}.
 \item $e_j$ is complex and $\Re{e_j}>0\,\forall j$: Trajectories spiral away from the fixed point. In this case the point is known as an 
\textit{unstable spiral}.
 \item $e_j$ is complex and $\Re{e_j}<0\,\forall j$: Trajectories spiral towards the fixed point; the point is known as a \textit{stable spiral} in 
this case.
\item There are a mixture of eigenvalues with differing signs for $\Re{e_j}$. In this case there are some stable directions and some unstable 
directions and the fixed point is known as a \textit{saddle point}.
\end{itemize}
At late times, trajectories in phase space flow away from unstable nodes and towards stable nodes. The system may spend prolonged periods of time 
near saddle points but will ultimately evolve towards a stable node. We have not discussed the case where the system contains one or more zero 
eigenvalues. The simplest cosmological systems all have non-zero eigenvalues but, as we have seen above, disformal theories contain a fixed point 
with two zero directions and so here we must use more advanced techniques. This is the subject of the next subsection.

\subsection{Centre Manifolds}\label{sec:appcm}

The centre manifold technique is used when the fixed point has one or more zero eigenvalues. Here we give a brief introduction and include only those 
aspects of the technique relevant for the problems studied in the main text. 

Let us consider a fixed point described by $n-p$ negative (or negative 
real part) eigenvalues with corresponding eigenvectors $\vec{E}_j$ and $p$ zero eigenvalues with eigenvectors $\vec{e}_j$\footnote{We are interested 
in stable fixed points since these describe the late-time behaviour of the system and so we will not consider positive eigenvalues here.}. One may 
expand the variables in terms of the eigenbasis such that
\begin{equation}
 X_i=\sum_{j=1}^{n-p} A_j^s (\vec{E}_j)_i+\sum_{j=n-p+1}^n A^c_j(\vec{e}_j)_i,
\end{equation}
where we label stable directions as $A_j^s$ and zero eigenvalue or \textit{centre} directions as $A^c_j$. The zero eigenvalues indicate that a linear 
analysis is not sufficient to analyse the stability in the directions of $\vec{e}_j$. This is a problem with two time scales. Trajectories in the 
directions of $\vec{E}_j$ will converge to the attractor swiftly but the behaviour of the components in the directions of $\vec{e}_j$ is unknown. The 
centre manifold technique allows one to solve for the late-time behaviour of $A_j^c$ by working on time scales such that $A_j^s$ have converged to 
their trajectories along the attractor. One then reduces the dimension of the phase space from $n$ to $p$ and can formulate the system as an 
autonomous one in the hope of being able to derive the behaviour of $A_j^c$. This works as follows: the full $n$-dimensional system can be described 
by the autonomous system
\begin{align}
 \frac{\dd A_i^s}{\dd t}&= g_i(\vec{A}^s,\vec{A}^c)\quad i=1,\ldots, n-p\quad\textrm{and}\\
 \frac{\dd A_i^c}{\dd t}&= h_i(\vec{A}^s,\vec{A}^c)\quad i= n-p+1,\ldots, n.
\end{align}
Setting, $g_i=0$ gives $n-p$ algebraic equations, which allows one to solve for $\vec{A}^s(\vec{A}^c)$ along trajectories at sufficiently late times 
such that the attractor has been reached in the $\vec{E}_j$ directions. The dynamics in the reduced, $p$-dimensional phase space are then described by
\begin{equation}
 \frac{\dd A_i^c}{\dd t}= h_i(\vec{A}^s(\vec{A}^c),\vec{A}^c)\quad i= n-p+1,\ldots, n.
\end{equation}
This represents a new Autonomous system that can be investigated using the techniques of Appendix \ref{sec:dyna} to discern the late-time behaviour.

\section{Eigenvalues at the Fixed Points}\label{sec:ape}
In this Appendix we list the eigenvalues for each fixed point. Points (18)--(20) are those of the two-dimensional phase space when $\gamma=\lambda/2$ 
and so there are only two eigenvalues.

\subsection{Eigenvalues}

\begin{widetext}
\begin{enumerate}
\item $e_1=\alpha ^2-{3}/{2}$, $e_2=-\alpha  (\alpha +2 \gamma )-{3}/{2}$, $e_3=\alpha  (\alpha +\lambda )+{3}/{2}$
 \item $e_1=3-\sqrt{6} \alpha$, $e_2=-\sqrt{6} \gamma -3$, $e_3=\sqrt{{3}/{2}} \lambda +3$
 \item $e_1=\sqrt{6} \alpha +3$, $e_2=\sqrt{6} \gamma -3$, $e_3=3-\sqrt{{3}/{2}} \lambda$
 \item $e_1= \lambda(\gamma-\lambda/2)$, $ 1/2(\lambda^2-6)$, $e_3=\lambda  (\alpha +\lambda )-3$
 \item $e_1={3 (2 \gamma -\lambda )}/{2 (\alpha +\lambda )}$, $e_2=-\frac{\sqrt{3} \sqrt{-(\alpha +\lambda )^2 \left(16 \alpha ^3 \lambda +4 \alpha ^2 
\left(8 \lambda ^2-15\right)+4 \alpha  \lambda  \left(4 \lambda ^2-9\right)+21 \lambda ^2-72\right)}+3 (\alpha +\lambda ) (2 \alpha +\lambda )}{4 
(\alpha +\lambda )^2}$,\\ $e_3=\frac{\sqrt{3} \sqrt{-(\alpha +\lambda )^2 \left(16 \alpha ^3 \lambda +4 \alpha ^2 \left(8 \lambda ^2-15\right)+4 
\alpha  \lambda  \left(4 \lambda ^2-9\right)+21 \lambda ^2-72\right)}-3 (\alpha +\lambda ) (2 \alpha +\lambda )}{4 (\alpha +\lambda )^2}$
 \item $e_1=-\frac{1}{2} \left(\sqrt{4 \gamma ^2-6}-2 \gamma \right) (2 \gamma -\lambda )$,\\ $e_2=\frac{1}{2} \left(3 \left(\sqrt{\frac{8 \alpha 
^2-4 \alpha  \left(\sqrt{4 \gamma ^2-6}-4 \gamma \right)+10 \gamma ^2-4 \gamma  \sqrt{4 \gamma ^2-6}-3}{2 \gamma  \left(\sqrt{4 \gamma ^2-6}+2 \gamma 
\right)-3}}-3\right)-\left(\sqrt{4 \gamma ^2-6}-2 \gamma \right) (2 \alpha +5 \gamma )\right)$,\\ $e_3=-\frac{1}{2} \left(\left(\sqrt{4 \gamma 
^2-6}-2 \gamma \right) (2 \alpha +5 \gamma )+3 \left(\sqrt{\frac{8 \alpha ^2-4 \alpha  \left(\sqrt{4 \gamma ^2-6}-4 \gamma \right)+10 \gamma ^2-4 
\gamma  \sqrt{4 \gamma ^2-6}-3}{2 \gamma  \left(\sqrt{4 \gamma ^2-6}+2 \gamma \right)-3}}+3\right)\right)$
 \item $e_1=\frac{1}{2} \left(\sqrt{4 \gamma ^2-6}+2 \gamma \right) (2 \gamma -\lambda )$, $e_2=\frac{3 \left(\left(\sqrt{4 \gamma ^2-6}-2 \gamma 
\right) 
 (2 \alpha -\gamma)+U_1-9\right)}{4 \gamma  \left(\sqrt{4 \gamma ^2-6}-2 \gamma \right)+6}$,\\ $e_3=-3\frac{U_2}{2 \left(\gamma  \left(\sqrt{4 \gamma 
^2-6}-2 \gamma \right)+3\right)^3}+\frac{3 \left(2 \alpha  \left(2 \gamma ^2-3\right) \left(\sqrt{4 \gamma ^2-6}-2 \gamma \right)+4 \gamma ^4-24 
\gamma ^2+3 \sqrt{4 \gamma ^2-6} \gamma -2 \sqrt{4 \gamma ^2-6} \gamma ^3+27\right)}{2 \left(\gamma  \left(\sqrt{4 \gamma ^2-6}-2 \gamma 
\right)+3\right)^2}$
\item $e_1=\sqrt{6} \gamma +3$, $e_2=-\sqrt{6} \alpha +\sqrt{6} \gamma +6$, $e_3=\sqrt{\frac{3}{2}} \lambda +3$
\item $e_1=3/2$, $e_2=3/2$, $e_3=3/2$
\item $e_1=3-\sqrt{6} \gamma$, $e_2=\sqrt{6} \alpha -\sqrt{6} \gamma +6$, $e_3=3-\sqrt{\frac{3}{2}} \lambda$
 \item $e_1=-3$, $e_2=0$, $e_3=0$
 \item $e_1=\frac{1}{9} \left(\left(\sqrt{4 (\alpha -\gamma )^2-18}+2 \alpha -2 \gamma \right) (\alpha -\gamma )-18\right)$, $e_2=\frac{1}{9} 
\left((\alpha +2 \gamma ) \left(\sqrt{4 (\alpha -\gamma )^2-18}+2 \alpha -2 \gamma \right)+9\right)$, $e_3=\frac{1}{18} \left(\left(\sqrt{4 (\alpha 
-\gamma )^2-18}+2 \alpha -2 \gamma \right) (2 \alpha -2 \gamma +3 \lambda )+18\right)$
 \item $e_1=\frac{1}{9} \left(\left(-\sqrt{4 (\alpha -\gamma )^2-18}+2 \alpha -2 \gamma \right) (\alpha -\gamma )-18\right)$, $e_2=\frac{1}{9} 
\left(9-(\alpha +2 \gamma ) \left(\sqrt{4 (\alpha -\gamma )^2-18}-2 \alpha +2 \gamma \right)\right)$, $e_3=\frac{1}{18} \left(18-\left(\sqrt{4 (\alpha 
-\gamma )^2-18}-2 \alpha +2 \gamma \right) (2 \alpha -2 \gamma +3 \lambda )\right)$
 \item $e_1=\frac{1}{2} \lambda  (\lambda -2 \gamma )$, $e_2=\frac{1}{2} \left(\lambda ^2-6\right)$, $e_3=\alpha  \lambda -\gamma  \lambda +\frac{3 
\lambda ^2}{2}-3$
 \item $e_1=\frac{3 \lambda -6 \gamma }{2 \alpha -2 \gamma +3 \lambda }$,\newline $e_2=-\frac{3 \left(U_3+(\alpha -\gamma +\lambda ) (\lambda  (2 
\alpha -2 \gamma +3 \lambda )-12) (2 \alpha -2 \gamma +3 \lambda )^{5/2}\right)}{(2 \alpha -2 \gamma +3 \lambda )^{7/2} (\lambda  (2 \alpha -2 \gamma 
+3 \lambda )-12)}$, $e_3=\frac{3 \left(U_3-(\alpha -\gamma +\lambda ) (2 \alpha -2 \gamma +3 \lambda )^{5/2} (\lambda  (2 \alpha -2 \gamma +3 
\lambda )-12)\right)}{(2 \alpha -2 \gamma +3 \lambda )^{7/2} (\lambda  (2 \alpha -2 \gamma +3 \lambda )-12)}$
 \item $e_1=\frac{1}{2} \left(\sqrt{4 \gamma ^2-6}+2 \gamma \right) (2 \gamma -\lambda )$,\newline $e_2=\frac{1}{2} \left(\left(\sqrt{4 \gamma 
^2-6}+2 \gamma \right) (2 \alpha +5 \gamma )+U_3-9\right)$, $e_3=\frac{1}{2} \left(\left(\sqrt{4 \gamma ^2-6}+2 \gamma \right) (2 \alpha +5 \gamma 
)-U_3-9\right)$
 \item $e_1=-\frac{1}{2} \left(\sqrt{4 \gamma ^2-6}-2 \gamma \right) (2 \gamma -\lambda )$,\newline$e_2=\frac{1}{2} \left(\left(\sqrt{4 \gamma 
^2-6}+2 \gamma \right) (2 \alpha +5 \gamma )+U_4-9\right)$, $e_3=\frac{1}{2} \left(\left(\sqrt{4 \gamma 
^2-6}+2 \gamma \right) (2 \alpha +5 \gamma )-U_4-9\right)$
  \item $e_1=\frac{1}{2} \left(\lambda ^2-6\right)$, $e_2=\lambda  (\alpha +\lambda )-3$
  \item $e_1=-\frac{6 \left(2 \alpha ^2 \lambda +3 \alpha  \left(\lambda ^2-4\right)+\lambda ^3\right)+\sqrt{3} U_4+3 \lt ^2 (2 
\alpha +\lambda ) (2 \alpha  (\alpha +\lambda )+3)-36 \lambda }{4 (\alpha +\lambda ) \left(2 (\alpha +\lambda ) \left(\alpha  \lt ^2+\lambda 
\right)+3 \left(\lt ^2-4\right)\right)}$,\newline $e_2=\frac{-6 \left(2 \alpha ^2 \lambda +3 \alpha  \left(\lambda ^2-4\right)+\lambda 
^3\right)+\sqrt{3} U_4-3 \lt ^2 (2 \alpha +\lambda ) (2 \alpha  (\alpha +\lambda )+3)+36 \lambda }{4 (\alpha +\lambda ) \left(2 (\alpha +\lambda 
) \left(\alpha  \lt ^2+\lambda \right)+3 \left(\lt ^2-4\right)\right)}$
\item $e_1=\frac{\left(\Lambda ^2-2\right) \left(\Lambda ^2 \left(4 \alpha  \lambda +5 \lambda ^2-18\right)-\Lambda  (4 \alpha +5 \lambda ) 
\sqrt{\left(\lambda ^2-6\right) \Lambda ^2+12}+36\right)-\sqrt{2}U_6 }{4 \left( \Lambda ^2-2\right) }$,\newline$e_2=\frac{\left(\Lambda ^2-2\right) 
\left(\Lambda ^2 \left(4 \alpha  \lambda +5 \lambda ^2-18\right)-\Lambda  (4 \alpha +5 \lambda ) 
\sqrt{\left(\lambda ^2-6\right) \Lambda ^2+12}+36\right)+\sqrt{2}U_6 }{4 \left( \Lambda ^2-2\right) }$
\end{enumerate}
where
\begin{align}U_1^2&=9-8 \alpha ^2 \left(2 \gamma  \left(\sqrt{4 \gamma ^2-6}-2 \gamma \right)+3\right)-4 \alpha  \left(4 \left(\sqrt{4 \gamma ^2-6}-2 
\gamma 
\right) \gamma ^2+3 \sqrt{4 \gamma ^2-6}\right)\nonumber\\&+8 \gamma ^4+6 \gamma ^2-6 \sqrt{4 \gamma ^2-6} \gamma -4 \sqrt{4 \gamma ^2-6} \gamma 
^3,\\U_2^2&=-\left(2 \gamma 
^2-3\right)^3 \left(8 \alpha ^2 \left(4 \gamma  \left(4 \gamma  \left(\gamma  \left(\sqrt{4 \gamma ^2-6}-2 \gamma \right)+3\right)-3 \sqrt{4 \gamma 
^2-6}\right)-9\right)\right.\nonumber\\&\left.+4 \alpha  \left(4 \gamma  \left(4 \left(2 \gamma  \left(\sqrt{4 \gamma ^2-6}-2 \gamma \right)+3\right) 
\gamma ^2+9\right)-9 
\sqrt{4 \gamma ^2-6}\right)-64 \gamma ^6+54 \gamma ^2+32 \sqrt{4 \gamma ^2-6} \gamma ^5+24 \sqrt{4 \gamma ^2-6} \gamma ^3+27\right)
\\U_3^2&=(2 \alpha -2 \gamma +3 \lambda )^5 
\left(38 \lambda ^3 (\alpha -\gamma )+\lambda ^2 \left(35 (\alpha -\gamma )^2-12\right)+10 \lambda  \left((\alpha -\gamma )^2-6\right) (\alpha 
-\gamma 
)\right.\nonumber\\&\left.-36 \left((\alpha -\gamma )^2+1\right)+12 \lambda ^4\right) (\lambda  (2 \alpha -2 \gamma +3 \lambda 
)-12),\\U_4^2&=\left(\left(\sqrt{4 \gamma ^2-6}+2 \gamma \right) (2 \alpha +5 \gamma )-9\right)^2-8 \left(\alpha  \left(4 \gamma  
\left(\gamma  \left(\sqrt{4 \gamma ^2-6}+2 \gamma \right)-3\right)-3 \sqrt{4 \gamma ^2-6}\right)\nonumber\right.\\&\left.+\gamma  \left(2 \gamma  
\left(4 \gamma  \left(\sqrt{4 
\gamma ^2-6}+2 \gamma \right)-15\right)-9 \sqrt{4 \gamma ^2-6}\right)+9\right),\\U_5^2&=-\left(\lt ^2 (2 \alpha  
(\alpha +\lambda )+3) \left(16 \alpha ^3 \lambda +4 \alpha ^2 \left(8 \lambda ^2-15\right)+4 \alpha  \lambda  \left(4 \lambda ^2-9\right)+21 \lambda 
^2-72\right)\nonumber\right.\\&\left.-6 \left(20 \alpha ^3 \lambda +8 \alpha ^2 \left(5 \lambda ^2-9\right)+3 \alpha  \left(7 \lambda ^2-16\right) 
\lambda +\lambda ^4+18 
\lambda ^2-72\right)\right) \left(2 (\alpha +\lambda ) \left(\alpha  \lt ^2+\lambda \right)+3 \left(\lt 
^2-4\right)\right)\quad\textrm{and}\\U_6^2&=\Lambda ^4 \left((4 \alpha +3 \lambda ) \left(4 \alpha  \left(\lambda 
^2-3\right)+3 \lambda  \left(\lambda ^2-5\right)\right)+18\right)-12 \Lambda  (4 \alpha +3 \lambda ) \sqrt{\left(\lambda ^2-6\right) \Lambda 
^2+12}\nonumber\\&-\Lambda ^3 (4 \alpha +3 \lambda ) \left(4 \alpha  \lambda +3 \lambda ^2-6\right) \sqrt{\left(\lambda ^2-6\right) 
\Lambda ^2+12}+6 \Lambda ^2 
((4 \alpha +3 \lambda ) (4 \alpha +5 \lambda )-12)+72.
\end{align}

\end{widetext}

\section{The $x$--$Y$--$W$ System}\label{sec:xYW}
Here, we present the autonomous system written using the variables $Y$ and $W$ defined in (\ref{eq:W}) and (\ref{eq:Y}):
\begin{widetext}
\begin{align}
 \frac{\dd x}{\dd N}&=\frac{\sqrt{6} \left(x^2+Y (Y+2)\right) \left((W+1) \left(2 \alpha +(W+1) \left(-\alpha +3 x \left(2 x (\alpha -\gamma 
)+\sqrt{6}\right)-3 \lambda  (Y+1)^2\right)\right)-\alpha \right)}{2(W+1) \left((W+1) \left(9 x^2+3 
(Y+1)^2-4\right)+2\right)-2}\nonumber\\&\label{eq:xYZ}+\frac{3}{2} x \left(x^2-Y (Y+2)\right)-3 x+\sqrt{6} \lambda  (Y+1)^2,\\
\frac{\dd Y}{\dd N}&=-\frac{1}{2} (Y+1) \left(-3 x^2+\sqrt{6} \lambda  x+3 Y (Y+2)\right)\label{eq:YYZ}\quad\textrm{and}\\\frac{\dd W}{\dd 
N}&=\frac{1}{2} W (W+1) \left(3 x^2-2 \sqrt{6} \gamma  x-3 Y (Y+2)\right)\label{eq:WYZ}.
\end{align}
\end{widetext}

\end{document}